\documentclass{article}

\usepackage[preprint,nonatbib]{neurips_2025}

\usepackage[utf8]{inputenc} 
\usepackage[T1]{fontenc}    
\usepackage{hyperref}       
\usepackage{url}            
\usepackage{booktabs}       
\usepackage{amsfonts}       
\usepackage{nicefrac}       
\usepackage{microtype}      
\usepackage{xcolor}         

\usepackage[pdftex]{graphicx}
\usepackage{amsthm}
\usepackage{amsmath}
\usepackage{amsfonts}
\usepackage{nccmath}
\usepackage{mathtools}
\usepackage{algorithm}
\usepackage{algorithmic}
\usepackage{comment}
\usepackage{url}
\usepackage{threeparttable}

\newcommand{\bE}{\mathbb E}

\newcommand{\bP}{\mathbb P}
\newcommand{\bR}{\mathbb R}

\newcommand{\cA}{\mathcal A}
\newcommand{\cB}{\mathcal B}

\newcommand{\cD}{\mathcal D}
\newcommand{\cE}{\mathcal E}
\newcommand{\cH}{\mathcal H}

\newcommand{\cN}{\mathcal N}
\newcommand{\cO}{\mathcal O}

\newcommand{\cR}{\mathcal R}

\newcommand{\cT}{\mathcal T}

\newcommand{\argmax}{\mathop{\rm argmax}\limits}

\newcommand{\E}{\mathop{\mathbb E}\limits}
\newcommand{\midd}{\,\middle|\,}

\newtheorem{theorem}{Theorem}
\newtheorem{definition}{Definition}
\newtheorem{lemma}{Lemma}
\newtheorem{corollary}{Corollary}
\newtheorem{proposition}{Proposition}

\title{Achieving PAC Guarantees in Mechanism Design through Multi-Armed Bandits
}

\author{%
  Takayuki Osogami\\
  IBM Research -- Tokyo\\
  \texttt{osogami@jp.ibm.com}\\
  \And
  Hirota Kinoshita\thanks{This research was conducted at IBM Research.}\\
  Max Planck Institute for Informatics\\
  \texttt{hirota.kinoshita@gmail.com}\\
  \And
  Segev Wasserkrug\\
  IBM Research -- Israel\\
  \texttt{segevw@il.ibm.com}
}

\begin{document}

\maketitle

\begin{abstract}
  We analytically derive a class of optimal solutions to a linear program (LP) for automated mechanism design that satisfies efficiency, incentive compatibility, strong budget balance (SBB), and individual rationality (IR), where SBB and IR are enforced in expectation. These solutions can be expressed using a set of essential variables whose cardinality is exponentially smaller than the total number of variables in the original formulation. However, evaluating a key term in the solutions requires exponentially many optimization steps as the number of players $N$ increases. We address this by translating the evaluation of this term into a multi-armed bandit (MAB) problem and develop a probably approximately correct (PAC) estimator with asymptotically optimal sample complexity. This MAB-based approach reduces the optimization complexity from exponential to $O(N\log N)$. Numerical experiments confirm that our method efficiently computes mechanisms with the target properties, scaling to problems with up to $N=128$ players---substantially improving over prior work.

\end{abstract}

\section{Introduction}
\label{sec:intro}

Mediators in multi-agent systems can improve efficiency by making centralized decisions that maximize social welfare. In trading networks, for example, they ensure goods are produced by the lowest-cost firms and allocated to those with the greatest need \cite{hatfield2013stability,osogami2023learning}. While such mediators could prioritize profit by charging participants---like today’s tech giants operating digital marketplaces or advertising platforms---they often capture most of the surplus, leaving participants with little gain.

We instead envision an open platform designed to maximize benefits for participants in multi-agent systems. This aligns with the goals of auction mechanisms for \emph{public} resources, where resources should go to those who value them most, and the mediator should run without budget deficits or surpluses \cite{bailey1997demand,cavallo2006optimal,dufton2021optimizing,gujar2011redistribution,guo2012worst,guo2009worst,manisha2018learning,tacchetti2022learning}.  However, such mechanisms rely on the structure of single-sided auctions, where all participants are buyers. In contrast, similar guarantees may be unattainable in more general settings---such as double-sided auctions \cite{hobbs2000evaluation,zou2009double,widmer2016efficiency,stoesser2010market,kumar2018truthful,chichin2017towards}, matching markets \cite{zhang2020optimal}, or trading networks \cite{osogami2023learning,wasserkrug2023benefits}---even when they are possible in the single-sided case.


We design mechanisms for general environments encompassing all the above multi-agent systems, focusing on efficiency and strong budget balance (SBB). Specifically, the mediator selects a social decision that maximizes the total value to the players (i.e., decision efficiency; DE), while ensuring its expected revenue equals a target $\rho \in \mathbb{R}$ (SBB when $\rho = 0$). Following standard Bayesian mechanism design \cite{shoham2009multiagent}, each player's valuation depends on their private type, while the joint distribution of types is common knowledge. Since DE is hard to achieve without knowing types, we require dominant strategy incentive compatibility (DSIC): truth-telling must be optimal regardless of others’ strategies. To promote participation, we also impose individual rationality (IR): each player's expected utility must exceed a type-dependent threshold $\theta(t_n) \in \mathbb{R}$ (with standard IR when $\theta \equiv 0$).

While these properties are standard in mechanism design \cite{shoham2009multiagent}, we introduce parameters, $\rho$ and $\theta$, to generalize the standard definitions of SBB and IR, motivated by three considerations. First, the standard definitions can make it impossible to satisfy all four desired properties simultaneously \cite{green1977characterization,myerson1983efficient,osogami2023learning}; our generalization enables precise characterization of when they are achievable. Second, it supports a principled, sample-based mechanism learning approach with theoretical guarantees, allowing the resulting mechanisms to approximately satisfy the four properties with high probability. Third, the parameters help model practical constraints---e.g., a mediator may require positive revenue to sustain the platform or need to guarantee players positive utility to attract participation.

We require SBB and IR to hold \emph{in expectation} (i.e., \emph{ex ante} or \emph{interim}) over the distribution of types, while DE and DSIC must hold for every realization of types (\emph{ex post}). These assumptions align with those in \cite{osogami2023learning,wasserkrug2023benefits} for trading networks, though the \emph{in expectation} requirements are clearly weaker than the \emph{ex post} guarantees typically assumed in auction settings \cite{bailey1997demand,cavallo2006optimal,dufton2021optimizing,gujar2011redistribution,guo2012worst,guo2009worst,manisha2018learning,tacchetti2022learning}. However, this relaxation allows us to derive \emph{analytical} solutions for mechanisms that satisfy all four desired properties in general environments and to characterize when such mechanisms exist. In contrast, prior analytical results are limited to auctions with a single good type \cite{bailey1997demand,cavallo2006optimal,guo2011vcg,guo2007worst,guo2009worst,moulin2009almost} or unit-demand settings \cite{gujar2011redistribution,guo2012worst}. For more complex auctions \cite{dufton2021optimizing,manisha2018learning,tacchetti2022learning} and trading networks \cite{osogami2023learning,wasserkrug2023benefits}, mechanisms are computed via numerical optimization, which often scales poorly with the number of players.

In particular, while the theoretical framework in \cite{osogami2023learning} could, in principle, be applied to any finite number of players, their numerical method is constrained by computational complexity and has been applied only to trading networks with \emph{two} players. In contrast, our analytical solutions can be evaluated numerically for around 10 players, depending on the number of types. The computational bottleneck in our analytical solutions is evaluating the minimum expected value over possible types for each player. Doing so exactly requires computing an efficient social decision for all $K^N$ type profiles, where $K$ is the number of possible types per player and $N$ is the number of players.

To address this bottleneck, we formulate the evaluation of the minimum expected value as a multi-armed bandit (MAB) problem \cite{lattimore2020bandit} and propose an asymptotically optimal algorithm. Unlike standard MAB objectives---regret minimization \cite{auer1995gambling,auer2002finite} or best arm identification \cite{audibert2010best,maron1993hoeffding,mnih2008empirical,bubeck2009pure}---our goal is to estimate the mean reward of the best arm. We develop a probably approximately correct (PAC) algorithm that estimates this value within $\varepsilon$ error with probability at least $1-\delta$, and show that its sample complexity, $O((K/\varepsilon^2)\,\log(1/\delta))$, matches our derived lower bound. This MAB-based approach reduces the number of computations of efficient social decisions from $K^N$ to $O(KN\log N)$, allowing us to handle cases up to $128$ players---substantial improvement over two players in \cite{osogami2023learning}.

Our contributions build on a previously formulated linear program (LP) for designing mechanisms that satisfy DE, SBB, DSIC, and IR, extending it with tunable parameters for SBB and IR (Section~\ref{sec:settings}-\ref{sec:LP}). While this extension is conceptually straightforward, it enables a more general analysis and a principled learning approach. First, we establish a sufficient condition for the feasibility of the LP and prove that it is also necessary when players’ types are independent (Section~\ref{sec:solution}). Second, under this condition, we analytically characterize a class of optimal solutions, yielding mechanisms that satisfy all four properties in general environments, including auctions and trading networks (Section~\ref{sec:solution}). Third, we formulate the evaluation of a key quantity in the analytical expressions as a best-mean estimation (BME) problem in MAB in order to provide a PAC guarantee to the four properties in the mechanism with minimal sample complexity (Section~\ref{sec:bandit}). Section~\ref{sec:bandit2} proposes an asymptotically optimal PAC algorithm for BME.  Finally, we empirically validate the effectiveness of our approach (Section~\ref{sec:exp}). We begin in Section~\ref{sec:prior} by positioning our contributions within prior work.

\section{Related work}
\label{sec:prior}

The most closely related prior work is \cite{osogami2023learning}, which formulates and numerically solves an LP to design mechanisms for trading networks that satisfy DE, DSIC, IR, and weak budget balance (WBB), where the mediator's expected revenue is nonnegative. While the LP in \cite{osogami2023learning} uses an arbitrary objective and includes SBB only as an example, our work focuses specifically on SBB and derives analytical solutions tailored to this objective. We also extend their formulation with tunable parameters.

The remainder of this section reviews related work on mechanism design---particularly efforts to achieve SBB---and on MAB, focusing on PAC algorithms. Notably, our approach is unique in estimating a key quantity from our analytically derived optimal solution, leading to the proposal of an asymptotically optimal PAC algorithm for the underexplored objective of best-mean estimation.

In single-sided auctions where only buyers act strategically, Vickrey–Clarke–Groves (VCG) mechanisms with Clarke’s pivot rule (VCG auctions) satisfy \textit{ex post} DE (often called allocative efficiency in auctions), DSIC, IR, and WBB \cite{nisan2007introduction}. However, the Green–Laffont Impossibility Theorem shows that no mechanism can achieve DE, DSIC, and SBB in all environments \cite{green1977characterization,green1979incentives}. This has led to efforts to redistribute the mediator’s revenue to players while maintaining DSIC, DE, IR, and WBB. Analytical mechanisms have been derived for auctions with single or homogeneous goods \cite{bailey1997demand,cavallo2006optimal,guo2011vcg,guo2007worst,guo2009worst,moulin2009almost}, or unit-demand bidders \cite{gujar2011redistribution,guo2012worst}. For auctions with multi-unit demands for heterogeneous goods, prior work has proposed numerical methods to approximate optimal redistribution, using piecewise linear functions \cite{dufton2021optimizing} or neural networks \cite{manisha2018learning,tacchetti2022learning}.

We consider general environments with heterogeneous goods and multi-unit demands, where players may have both negative and positive valuations for social decisions (e.g., players may buy or sell goods depending on the social decision). Under these conditions, the Myerson-Satterthwaite Impossibility Theorem \cite{myerson1983efficient} shows that no mechanism can guarantee \emph{ex post} DE, DSIC, IR, and WBB in all such environments, unlike VCG auctions. Thus, we derive mechanisms that achieve DE, DSIC, IR, and SBB as best as possible. A limitation of our results is that IR and SBB hold only in expectation, but this is justifiable for risk-neutral mediators and players \cite{osogami2023learning}. Additionally, our model can guarantee \emph{strictly} positive expected utility, which can ensure nonnegative utility with high probability when players repeatedly participate in the mechanism.

Much of the literature focuses on maximizing the mediator's revenue in auctions \cite{myerson1981optimal}, with recent work on automated mechanism design (AMD) using machine learning \cite{duetting2019optimal,rahme2021auction,ivanov2022optimaler,curry2020certifying} and analyzing sample complexity \cite{balcan2016sample,morgenstern2015pseudo,syrgkanis2017sample}. Similar to these studies, we formulate an optimization problem whose solution yields a mechanism with the desired properties. However, rather than solving it numerically, we derive optimal solutions analytically. While prior work analyzes the sample complexity for maximizing expected revenue, we focus on the sample complexity of evaluating specific expressions in analytically derived mechanisms.

We evaluate our analytical expression using best-mean estimation (BME) in MAB, where standard objectives include regret minimization \cite{auer1995gambling,auer2002finite} and best arm identification (BAI) \cite{audibert2010best,maron1993hoeffding,mnih2008empirical,bubeck2009pure}. The most relevant prior work on MAB is PAC learning for BAI and its sample complexity analysis. We reduce BME to BAI and establish a lower bound on BME's sample complexity using a technique from BAI \cite{evendar02pac}.  While this method does not give tight bounds for BAI \cite{mannor2004sample}, it provides tight bounds for BME.  Note that the problem of estimating the best mean frequently appears in reinforcement learning \cite{hasselt2010double} and machine learning \cite{kajino2023biases}, where the focus is on estimating the best mean from a given set of samples \cite{hasselt2013estimating}, while our focus is on efficiently collecting samples for the estimation.

\section{Settings}
\label{sec:settings}

The goal of mechanism design is to specify the rules of a game in a way that an outcome desired by the mechanism designer is achieved when rational players, aiming to maximize their individual utility, participate \cite{jackson2014mechanism,shoham2009multiagent}.  Let $\cN\coloneqq[1,2,\ldots,N]$ be the set of players and $\cO$ be the set of possible outcomes.  For each player $n\in\cN$, let $\cA_n$ be the set of available actions and $\cT_n$ be the set of possible types.  Let $\cA\coloneqq\cA_1\times\ldots\times\cA_N$ and $\cT\coloneqq\cT_1\times\ldots\times\cT_N$ be the corresponding product spaces.  A mechanism $\mu:\cA\to\cO$ determines an outcome depending on the actions taken by the players.  Let $u_n:\cO\times\cT_n\to\bR$ be the utility function of each player $n\in\cN$.

We consider Bayesian games where the players' types follow a probability distribution that is known to all players and the mediator.  Before selecting actions, the players know their own types but not the types of the other players.  A strategy of each player $n\in\cN$ is thus a function from $\cT_n$ to $\cA_n$.

We assume that an outcome is determined by a social decision and payment; hence, a mechanism $\mu$ consists of a decision rule and a payment rule.  Let $\cD$ be the set of possible social decisions.  Given the actions of the players, the decision rule $\phi:\cA\to\cD$ determines a social decision, and the payment rule $\tau:\cA\to\bR^\cN$ determines the amount of (possibly negative) payment to the mediator from each player.  Let $v:\cD\times(\cT_1\cup\ldots\cup\cT_N)\to\bR$ specify the value of a given social decision to the player of a given type.  Then the utility of player $i$ when players take actions $a\in\cA$ is
\begin{align}
    u_n(\mu(a); t_n)
    = u_n((\phi(a), \tau(a)); t_n)
    = v(\phi(a); t_n) - \tau_n(a).\notag
\end{align}
We assume that $\cN$, $\cD$, and $\cT_n, \forall n\in\cN$ are finite sets.

Without loss of generality by the revelation principle \cite{shoham2009multiagent}, we consider only direct mechanisms, where the action available to each player is to declare their type to from the set of possible types (i.e., $\cA_n=\cT_n, \forall n\in\cN$).  We thus use $\cT_n$ for $\cA_n$.

Then we seek to achieve the following four properties with our mechanisms.  The first property is Dominant Strategy Incentive Compatibility (DSIC), which ensures that the optimal strategy of each player is to truthfully reveal its type
regardless of the strategies of the other players.  Formally,
\begin{fleqn}[10pt]
\begin{align}
    \mbox{[DSIC]} \quad
    v(\phi(t_n,t_{-n}'); t_n) - \tau_n(t_n,t_{-n}')
    \ge v(\phi(t'); t_n) - \tau_n(t'),
    \forall t'\in\cT, \forall t_n\in\cT_n, \forall n\in\cN,
    \notag
\end{align}
\end{fleqn}
where the left-hand side represents the utility of the player having type $t_n$ when it declares the same $t_n$, and the other players declare arbitrary types $t_{-n}'$.

The second property is Decision Efficiency (DE), which requires that the mediator chooses the social decision that maximizes the total value to the players.  With DSIC, we can assume that the players declare true types, and hence we can write DE as a condition on the decision rule:
\begin{fleqn}[10pt]
\begin{align}
    \mbox{[DE]} \qquad
    \phi(t) \in \argmax_{d\in\cD} \sum_{n\in\cN} v(d; t_n)
    \qquad\forall t\in\cT.
    \notag
\end{align}
\end{fleqn}

As the third property, we generalize individual rationality and require that the expected utility of each player is at least as large as a target value that can depend on its type.  We refer to this property as $\theta$-IR.  Again, assuming that players declare true types due to DSIC, we can write $\theta$-IR as follows:
\begin{fleqn}[10pt]
  \begin{align}
    \mbox{[$\theta$-IR]} \quad
    \bE[ v(\phi(t); t_n) - \tau_n(t) \mid t_n ]
    \ge \theta(t_n)
    \qquad \forall t_n\in\cT_n, \forall n\in\cN,
    \notag
\end{align}
\end{fleqn}
where $\theta:\cT_1\cup\ldots\cup\cT_N\to\bR$ determines the target expected utility for each type.  Throughout (except in Section~\ref{sec:bandit}, where we discuss general MAB models), $\bE$ denotes the expectation with respect to the probability distribution $\bP$ of types, which is the only probability that appears in our mechanisms.

The last property is a generalization of Budget Balance (BB), which we refer to as $\rho$-WBB and $\rho$-SBB.  Specifically, $\rho$-WBB requires that the expected revenue of the mediator is no less than a given constant $\rho\in\bR$, and $\rho$-SBB requires that it is equal to $\rho$.  Again, assuming that the players declare true types due to DSIC, these properties can be written as follows:
\begin{fleqn}[10pt]
\begin{align}
    \mbox{[$\rho$-WBB]} \qquad
    \sum_{n\in\cN} \bE\left[ \tau_n(t) \right] \ge \rho.
    \qquad\qquad\qquad\qquad\quad
    \mbox{[$\rho$-SBB]} \qquad
    \sum_{n\in\cN} \bE\left[ \tau_n(t) \right] = \rho.
    \notag
\end{align}
\end{fleqn}
While $\rho$-SBB is stronger than $\rho$-WBB, we will see that $\rho$-SBB is satisfiable if and only if $\rho$-WBB is satisfiable.

\section{Optimization problem for automated mechanism design}
\label{sec:LP}

Following \cite{osogami2023learning}, we seek to find optimal mechanisms in the class of VCG mechanisms, specified by a pair $(\phi^\star,h)$.  Specifically, after letting player take the actions of declaring their types $t\in\cT$, the mechanism first finds a social decision $\phi^\star(t)$ using a decision rule $\phi=\phi^\star$ that satisfies DE.  It then determines the amount of payment from each player $n\in\cN$ to the mediator by
\begin{align}
    \tau_n(t)
    & = h_n(t_{-n}) - \sum_{m\in\cN_{-n}} v(\phi^\star(t); t_m),
    \label{eq:VCG-payment}
\end{align}
where we define $\cN_{-n}\coloneqq\cN\setminus\{n\}$, and $h_n:\cT_{-n}\to\bR$ is an arbitrary function of the types of the players other than $n$ and referred to as a pivot rule.  The decision rule $\phi^\star$ guarantees DE by construction, and the payment rule \eqref{eq:VCG-payment} then guarantees DSIC \cite{nisan2007introduction}.

Our problem is now reduced to find the pivot rule, $h=\{h_n\}_{n\in\cN}$, that minimizes the expected revenue of the mediator, while satisfying $\theta$-IR and $\rho$-WBB.  This may lead to satisfying $\rho$-SBB if the revenue is maximally reduced.  To represent this reduced problem, let
\begin{align}
  w^\star(t) & \coloneqq \sum_{n\in\cN} v(\phi^\star(t); t_n)
  \label{eq:wstar}
\end{align}
be the total value of the efficient social decision when the players have types $t$.  Then we can rewrite $\theta$-IR (for the player having type $t_n$) and $\rho$-WBB as follows (see Appendix~\ref{sec:detail:LP} for full derivation):
\begin{align}
    \bE[ v(\phi^\star(t); t_n) - \tau_n(t) \mid t_n ]
    \ge \theta(t_n)
    & \Longleftrightarrow 
    \bE\left[ w^\star(t) \midd t_n \right]
    - \bE\left[ h_n(t_{-n}) \midd t_n \right]
    \ge \theta(t_n)
    \label{eq:IRequiv} \\
    \sum_{n\in\cN} \bE\left[ \tau_n(t) \right] \ge \rho
    & \Longleftrightarrow
    \sum_{n\in\cN} \bE\left[ h_n(t_{-n})\right]
    - (N-1) \, \bE\left[ w^\star(t) \right] \ge \rho.
    \label{eq:WBBequiv}
\end{align}

Therefore, we arrive at the following linear program (LP):
\begin{align}
    \min_h
    & \qquad \sum_{n\in\cN} \mathbb{E}[h_n(t_{-n})]
    \label{eq:LP-obj}\\
    \mathrm{s.t.}
    & \qquad \bE\left[ w^\star(t) \midd t_n \right]
    - \bE\left[ h_n(t_{-n}) \midd t_n \right]
    \ge \theta(t_n)
    \qquad \forall t_n\in\cT_n, \forall n\in\cN
    \label{eq:LP-IR} \\
    & \qquad \sum_{n\in\cN} \bE\left[ h_n(t_{-n})\right]
    - (N-1) \, \bE\left[ w^\star(t) \right] \ge \rho.
    \label{eq:LP-WBB}
\end{align}
The approach in \cite{osogami2023learning} solves this LP numerically (considering only the case with $\rho = 0$ and $\theta \equiv 0$). Since there is one variable $h_n(t_{-n})$ for each $t_{-n} \in \cT_{-n}$ and $n \in \cN$, the LP involves $N\,K^{N-1}$ variables and $N\,K + 1$ constraints when each player has $K$ possible types. If the LP is feasible, let $h^\star$ denote its optimal solution; the resulting VCG mechanism $(\phi^\star, h^\star)$ then satisfies DSIC, DE, $\theta$-IR, and $\rho$-WBB (Proposition~\ref{prop:LP} in Appendix~\ref{sec:detail:LP}).  Otherwise, no such VCG mechanism exists. In Section~\ref{sec:solution}, we fully characterize the feasibility condition and derive analytical solutions.

\section{Analytical solutions to the optimization problem}
\label{sec:solution}


We first establish a sufficient condition and a necessary condition for the LP to have feasible solutions.
\begin{lemma}
  The LP given by \eqref{eq:LP-obj}-\eqref{eq:LP-WBB} is feasible if
  \begin{align}
    \sum_{n\in\cN} \min_{t_n\in\cT_n} \left\{
      \bE[w^\star(t) \mid t_n] - \theta(t_n)
    \right\}
    & \ge (N - 1) \, \bE[w^\star(t)] + \rho.
    \label{eq:condition}
  \end{align}
  When types are independent ($t_m$ and $t_n$ are independent for any $m\neq n$ under $\bP$),
  the LP feasible if and only if \eqref{eq:condition} holds.
  When types are dependent, the LP may be feasible even if \eqref{eq:condition} is violated.
  \label{lemma:condition}
\end{lemma}
Complete proofs of all theoretical statements are provided in Appendix~\ref{sec:proof}. The proof of Lemma~\ref{lemma:condition} offers intuition behind the condition~\eqref{eq:condition}, which arises as the requirement for a constant pivot rule to satisfy $\theta$-IR and $\rho$-WBB. A key insight of our results is that this sufficient condition is also necessary when player types are independent.

Building on this, we analytically derive a class of \emph{optimal} solutions to the LP under condition~\eqref{eq:condition}, regardless of whether types are independent or not:
\begin{lemma}
  \label{lemma:constant}
  A pivot rule is called constant if and only if
  there exists a constant $\eta_n$
  such that $h_n(t_{-n})=\eta_n, \forall t_{-n}\in\cT_{-n}$ for each $n\in\cN$. Let $\cH$ be the set of constant pivot rules with:
  \begin{align}
    \eta_n
    & = \min_{t_n\in\cT_n} \left\{
      \bE[w^\star(t) \mid t_n]  - \theta(t_n)
    \right\} - \delta_n
    \qquad \forall n\in\cN \label{eq:eta-n}
  \end{align}
  where $\delta$ lies on the following simplex:
  \begin{align}
    \delta_n
    & \ge 0
    \qquad \forall n\in\cN \label{eq:simplex1}\\
    \sum_{n\in\cN} \delta_n
    & = \sum_{n\in\cN} \min_{t_n\in\cT_n}
    \left\{
      \bE[w^\star(t) \mid t_n]  - \theta(t_n)
    \right\}
    - (N - 1) \, \bE[w^\star(t)] - \rho. \label{eq:simplex2}
  \end{align}
  Then, when \eqref{eq:condition} holds, $\cH$ is nonempty, and any $h\in\cH$ is an optimal solution to the LP \eqref{eq:LP-obj}-\eqref{eq:LP-WBB}.
  Also, there exists a pivot rule in $\cH$ that gives a feasible solution to the LP if and only if \eqref{eq:condition} holds.
  \label{lemma:optimal}
\end{lemma}
In deriving the optimal solutions, we have substantially reduced the essential number of variables (from $N\,T^{N-1}$ to $N$ when each player has $T$ types).  Our approach can therefore not only find but also represent or store solutions exponentially more efficiently than \cite{osogami2023learning}.  Moreover, it turns out that the solutions in $\cH$ not only satisfy $\rho$-WBB but also $\rho$-SBB (in addition to DE, DSIC, and $\theta$-IR) regardless of whether the types are independent or not (formally proved as Corollary~\ref{cor:BB} in Appendix~\ref{sec:proof}).

When types are independent, the condition \eqref{eq:condition} is necessary for the existence of a feasible solution; hence, we do not lose optimality by considering only the solutions in $\cH$.  When types are dependent, the condition \eqref{eq:condition} may still be satisfied, and \emph{the solutions in $\cH$ remain optimal in this case}.  However, when types are dependent,
the LP may be feasible even if \eqref{eq:condition} is violated, and in this case optimal solutions are not in the space of constant pivot rules.

When the LP is infeasible, we may construct a mechanism that satisfies one of $\rho$-SBB and $\theta$-IR (in addition to DE and DSIC) regardless of whether the types are independent or not.  Specifically, any VCG mechanism with a pivot rule that satisfies \eqref{eq:eta-n} and \eqref{eq:simplex2} ensures $\rho$-SBB, and any VCG mechanism with a pivot rule that satisfies \eqref{eq:eta-n} and \eqref{eq:simplex1} ensures $\theta$-IR for any $t_n\in\cT_n$ and $n\in\cN$.  These are formally proved as Corollary~\ref{cor:SBB-IR} in Appendix~\ref{sec:proof}.  For example, for any $t_n\in\cT_n$ and $n\in\cN$, the following pivot rule always satisfies $\theta$-IR:
\begin{align}
  \eta_n
  & = \min_{t_n\in\cT_n} \left\{
    \bE[w^\star(t) \mid t_n]  - \theta(t_n)
  \right\} - \max\{\delta, 0\} \qquad \forall n\in\cN,
  \label{eq:optimal-eta-IR}
\end{align}
where we define (see also \eqref{eq:optimal-delta} in Appendix~\ref{sec:proof})
\begin{align}
  \delta
  & \coloneqq \frac{1}{N} \bigg(
    \sum_{n\in\cN} \min_{t_n\in\cT_n} \left\{
      \bE[w^\star(t) \mid t_n]  - \theta(t_n)
    \right\}
    - (N - 1) \, \bE[w^\star(t)] - \rho
  \bigg).
  \label{eq:optimal-delta-main}
\end{align}

Alternatively, one may choose $\theta$ and $\rho$ in a way that they ensure feasibility of the LP (i.e., \eqref{eq:condition} is satisfied).  For example, \eqref{eq:condition} is satisfied if we set $\theta \equiv 0$ and
\begin{align}
  \rho
  & = \left[
    \sum_{n\in\cN} \min_{t_n\in\cT_n} \bE[w^\star(t)\mid t_n]
    - (N-1) \, \bE[w^\star(t)]
  \right]^-,
  \label{eq:solution:rho-theta}
\end{align}
where $[x]^-\coloneqq\min\{x,0\}$ for $x\in\bR$.  When $\rho<0$, the mediator might get negative expected revenue, but the expected loss of the mediator is at most $|\rho|$.  Appendix~\ref{sec:detail:solution:WBB} provides another choice of $\theta$ with $\rho = 0$ that also guarantees feasibility, but may cause some players to incur negative expected utility.

While our analytical solutions significantly reduce computational cost compared to the numerical approach in \cite{osogami2023learning}, they still require evaluating
\begin{align}
  \kappa_n(\theta)
  & \coloneqq \min_{t_n\in\cT_n} \left\{
    \bE[w^\star(t)\mid t_n] -\theta(t_n)
  \right\}
  \label{eq:min-heavy}
\end{align}
for each $n\in\cN$.  Since $\bE$ is the expectation over the distribution $\bP$ on $\cT$, this requires evaluating $w^\star(t)$ for all $t\in\cT$.  From \eqref{eq:wstar}, $w^\star(t)$ is the total value of the efficient decision $\phi^\star(t)$, which itself is the solution to an optimization problem defining DE.  Without any structure in $\cD$ or $v$, this requires evaluating the total value for every decision in $\cD$.

\section{Evaluating the analytical solutions with a PAC guarantee}
\label{sec:bandit}

To reduce the computational cost of \eqref{eq:min-heavy}, we adopt a learning-based approach. The key observation is that estimating \eqref{eq:min-heavy} can be cast as a variant of a multi-armed bandit (MAB) problem, where the goal is to estimate the mean reward of the best arm.  Specifically, pulling an arm $t_n\in\cT_n$ yields a reward of $\theta(t_n) - w^\star(t)$, with $t$ sampled from the conditional distribution $\bP[\cdot\mid t_n]$.  To be consistent with prior MAB literature \cite{evendar02pac,evendar06action,hassidim2020optimal,mannor2004sample}, we frame the problem as one of reward maximization.


Since we assume that $\cN$, $\cD$, and $\cT_n, \forall n\in\cN$ are finite, there exist constants, $\bar\theta$ and $\bar v$, such that $|\theta(t')|\le\bar\theta$ and $|v(d;t')|\le \bar v, \forall d\in\cD, \forall t'\in\cup_{n\in\cN} \cT_n$.  Then we can also bound $|\theta(t_n) - w^\star(t)|\le \bar\theta + N\,\bar v, \forall t\in\cT, \forall n\in\cN$.  Namely, the reward is bounded.  We assume that the bounds are known, allowing us to scale the reward to the interval $[0,1]$ surely.

We also assume that we have access to an arbitrary size of the sample that is independent and identically distributed (i.i.d.) according to $\bP[\cdot\mid t_n]$ for any $t_n\in\cT_n, n\in\cN$.  When players have independent types, such sample can be easily constructed as long as we have access to i.i.d.\ sample $\{t^{(i)}\}_{i=1,2,\ldots}$ from $\cT$, because $\{(t_n,t_{-n}^{(i)})\}_i$ is the sample from $\bP[\cdot\mid t_n]$ for any $t_n\in\cT_n, n\in\cN$.


Consider the general $K$-armed bandit where each arm's reward is bounded in $[0,1]$.  For each $k\in[1,2,\ldots,K]$, let $\mu_k$ be the true mean of arm $k$.  Let $\mu_\star\coloneqq\max_k\mu_k$ be the best mean-reward, which we seek to estimate.  We say that the sample complexity of an algorithm for a MAB is $T$ if the algorithm pulls arms at most $T$ times.

A standard PAC algorithm for MAB returns an $\varepsilon$-optimal arm with probability at least $1-\delta$ for given $\varepsilon, \delta$ \cite{evendar06action,hassidim2020optimal}.  On the other hand, we need to evaluate \eqref{eq:min-heavy} within a given estimation error with high probability.
Formally, we will use the following definitions:
\begin{definition}
  For $\varepsilon,\delta>0$, we say that an algorithm is ($\varepsilon,\delta$)-PAC Best Arm Identifier (BAI) if the output $\hat I$ returned by the algorithm satisfies
  $\Pr\left( \mu_{\hat I} \ge \mu_\star - \varepsilon \right) \ge 1 - \delta$ and that
    an algorithm is ($\varepsilon,\delta$)-PAC Best Mean Estimator (BME) if the output $\hat\mu$ returned by the algorithm satisfies
    $\Pr\left( |\hat\mu - \mu_\star| \le \varepsilon \right) \ge 1 - \delta$.
    \label{def:PAC-BAI-BME}
\end{definition}

BAI and BME are related but different.  For example, consider a case where the best arm has large variance and $\mu_\star=1/2$, and all the other arms always yield zero reward $\mu_n=0, \forall n\neq\star$.  Then BAI is relatively easy here due to the large gap $\mu_\star-\mu_n=1/2, \forall n\neq\star$, while BME would demand more samples to accurately estimate $\mu_\star$ due to the large variance of the best arm.
In contrast, suppose there are many arms with Bernoulli rewards: half have expected value 1, and the rest have expected value $1-(3/2)\,\varepsilon$.  By pulling sufficiently many arms (once for each arm), Hoeffding’s inequality allows us to estimate that the best mean is at least $1-\varepsilon$ with high probability---sufficient for BME.  However, BAI would require identifying a specific arm with high expected value, necessitating many samples to distinguish it reliably (to be able to say that this particular arm has expected value at least $1-\varepsilon$).

We can estimate the term $\kappa_n(\theta)$ in \eqref{eq:min-heavy} by the use of an $(\varepsilon,\delta)$-PAC BME.  The optimal solutions in Lemma~\ref{lemma:optimal} involves another term,
\begin{align}
  \lambda(\rho)
    & \coloneqq \E[w^\star(t)] + \rho / (N-1),
\end{align}
with expectation, but this can also be estimated using a standard PAC estimator for expectation.
These allow us to find a mechanism that satisfies designed properties with high probability.  Formally, recalling that $N$ is the number of players, we have the following lemma:
\begin{lemma}
    Let $\tilde\kappa_n(\theta)$ for $n\in\cN$ and $\tilde\lambda(\rho)$ be independent estimates of $\kappa_n(\theta)$ and $\lambda(\rho)$ respectively given by an $(\varepsilon',\delta')$-PAC Best Mean Estimator and a standard $(\varepsilon'',\delta')$-PAC estimator of expectation. Also, let $\tilde d \coloneqq d(\tilde\kappa(\theta),\tilde\lambda(\rho),\varepsilon''',\varepsilon'''')$ be a point on the following simplex (here, we change the notation from $\delta$ in Lemma~\ref{lemma:optimal} to $\tilde d$ to avoid confusion):
    \begin{align}
        \tilde d_n
        & \ge \varepsilon''', \forall n\in\cN \\
        \sum_{n\in\cN}\tilde d_n
        & = \sum_{n\in\cN} \tilde\kappa_n(\theta) - (N-1) \, (\tilde\lambda(\theta)+\varepsilon'''')
    \end{align}
    Then the VCG mechanism with the constant pivot rule $h_n(t_{-n})=\eta_n=\tilde\kappa_n(\theta)-\tilde d_n$ satisfies $(\theta-(\varepsilon'-\varepsilon'''))$-IR and $(\rho-(N-1)\,(\varepsilon''-\varepsilon''''))$-WBB with probability $(1-\delta')^{N+1}$.
    \label{lemma:connect}
\end{lemma}

Notice that the sufficient condition of Lemma 1 states that the simplex in Lemma~\ref{lemma:optimal} is nonempty. Analogously, when the simplex in Lemma~\ref{lemma:connect} is empty, we cannot provide the solution that guarantees the properties stated in Lemma~\ref{lemma:connect}. Since DSIC and DE remain satisfied regardless of whether $\kappa_n$ for $n\in\cN$ and $\lambda$ are estimated or exactly computed, Lemma~\ref{lemma:connect} immediately establishes the following theorem by appropriately choosing the parameters:
\begin{theorem}
    In Lemma~\ref{lemma:connect}, let $\varepsilon'''=\varepsilon'$, $\varepsilon''''=\varepsilon''$, and $\delta'=1 - (1-\delta)^{1/(N+1)}$.  Then the VCG mechanism with the constant pivot rule $h_n(t_{-n})=\tilde\kappa_n(\theta)-\tilde d_n$ satisfies DSIC, DE, $\theta$-IR, and $\rho$-WBB with probability $1-\delta$.
    \label{thrm:connect}
\end{theorem}

Recall that computing our mechanism from Lemma~\ref{lemma:optimal} requires evaluating $w^\star(t)$ for all $t\in\cT$, which grows exponentially with the number of players $N$.  In Section~\ref{sec:bandit2}, we show that there exists an $(\varepsilon,\delta)$-PAC BME with sample complexity $O((K/\varepsilon^2) \, \log(1/\delta))$, which can be used to reduce the number of required evaluation of $w^\star(t)$ to $O(N\,\log N)$, as is formally proved in the following proposition:
\begin{proposition}
    The sample complexity to learn the constant pivot rule in Theorem~\ref{thrm:connect} is $O((N\,K/\varepsilon^2)\,\log(N/\delta))$, where $N=|\cN|$ is the number of players, and $K=\max_{n\in\cN} \cT_n$ is the maximum number of possible types of each player.
    \label{prop:connect}
\end{proposition}

\section{Estimating the best mean reward in multi-armed bandits}
\label{sec:bandit2}

Here, we establish matching lower and upper bounds on the sample complexity for ($\varepsilon,\delta$)-PAC BME:
\begin{theorem}
    There exists an ($\varepsilon,\delta$)-PAC BME with sample complexity $O((K/\varepsilon^2)\,\log(1/\delta))$, and any ($\varepsilon,\delta$)-PAC BME must have the sample complexity at least $\Omega((K/\varepsilon^2)\,\log(1/\delta))$.
    \label{thrm:complexity-bme}
\end{theorem}
These bounds are match those for ($\varepsilon,\delta$)-PAC BAI \cite{mannor2004sample}, indicating that BME and BAI share similar sample complexity, despite their differences discussed in Section~\ref{sec:bandit}.



Our upper bound is established by reducing BME to BAI.  Suppose that we have access to an arbitrary $(\varepsilon,\delta)$-PAC BAI with sample complexity $M$.  We can construct a $((3/2)\varepsilon,2\delta)$-PAC BME by first running the $(\varepsilon,\delta)$-PAC BAI and then taking $m^\star$ samples from the arm $\hat I$ that is identified as the best to estimate its mean (Algorithm~\ref{alg:best-reward} in Appendix~\ref{sec:detail:bandit2}).  When $m^\star$ is appropriately selected, the following lemma holds:
\begin{lemma}
    When an $(\varepsilon,\delta)$-PAC BAI with sample complexity $M$ is used, Algorithm~\ref{alg:best-reward} has sample complexity $M+m^\star$, where 
    $m^\star\coloneqq 
    \left\lceil
    (2/\varepsilon^2) \log (1.22/\delta)
    \right\rceil$,
    and returns $\hat\mu_{\hat I}$ with
    \begin{align}
        \Pr\left( |\hat\mu_{\hat I} - \mu_\star| > (3/2)\varepsilon \right) \le 2 \, \delta.
        \label{eq:toshow-upperbound}
    \end{align}
    \label{lemma:bai2bme}
\end{lemma}
Since there exists an ($\varepsilon,\delta$)-PAC BAI with sample complexity $O((K/\varepsilon^2)\,\log(1/\delta))$ \cite{mannor2004sample}, this establishes the upper bound on the sample complexity in Theorem~\ref{thrm:complexity-bme}.


Our lower bound is established by showing that an arbitrary $(\varepsilon,\delta)$-PAC BME can be used to solve the problem of identifying whether a given coin is negatively or positively biased (precisely, $\varepsilon$-Biased Coin Problem of Definition~\ref{def:bias} in Appendix~\ref{sec:detail:bandit2}), for which any algorithm is known to require \emph{expected} sample complexity at least $\Omega((1/\varepsilon^2)\,\log(1/\delta))$ to solve correctly with probability at least $1-\delta$ \cite{chernoff1972sequential,evendar02pac} (see Lemma~\ref{lemma:bias} in Appendix~\ref{sec:detail:bandit2}).  The following lemma, together with Lemma~\ref{lemma:bias}, establishes the lower bound on the sample complexity in Theorem~\ref{thrm:complexity-bme}:
\begin{lemma}
    If there exists an $(\varepsilon/2,\delta/2)$-PAC BME with sample complexity $M$ for $K$-armed bandit, then there also exists an algorithm, having expected sample complexity $M/K$, that can solve the $\varepsilon$-Biased Coin Problem correctly with probability at least $1-\delta$.
    \label{lemma:lower}
\end{lemma}

The proof technique of reduction from the $\varepsilon$-Biased Coin Problem was also used in \cite{evendar02pac} to prove a lower bound on the sample complexity of BAI.  What is interesting, however, is that the lower bound in \cite{evendar02pac} is not tight when $\delta<1/K$, and a tight lower bound is later established by a different technique in \cite{mannor2004sample}.  In contrast, our reduction gives a tight lower bound on the sample complexity of BME.  Appendix~\ref{sec:detail:bandit2} further explains the source of this difference.


\section{Numerical experiments}
\label{sec:exp}

We conduct numerical experiments to validate our approach and explore its limitations, focusing on two key questions:
\romannumeral 1) How effectively does BME reduce the number of evaluations of $w^\star(t)$?  \romannumeral 2) How well do $\theta$-IR and $\rho$-SBB hold when the $\kappa_n(\theta)$ in \eqref{eq:min-heavy} is estimated via BME?  While full details are in Appendix~\ref{sec:detail:exp}, our results show that BME substantially reduces computational complexity---by several orders of magnitude---for a moderate ($\sim 10$) to large number of players, though its effectiveness diminishes as the number of types increases.  Throughout our experiments, we use Successive Elimination for BME (SE-BME; Algorithm~\ref{alg:SE-BME} in Appendix~\ref{sec:detail}) as the implementation of BME.  All experiments were conducted in a cloud environment using a single CPU core and up to 66~GB of memory, without GPU acceleration (see Appendix~\ref{sec:detail:exp:require} for details).  The source code is included as a supplementary material and will be open-sourced upon acceptance.




\begin{figure}[t]
    \centering
    \begin{minipage}{0.3\linewidth}
    \centering
     \includegraphics[width=0.95\linewidth]{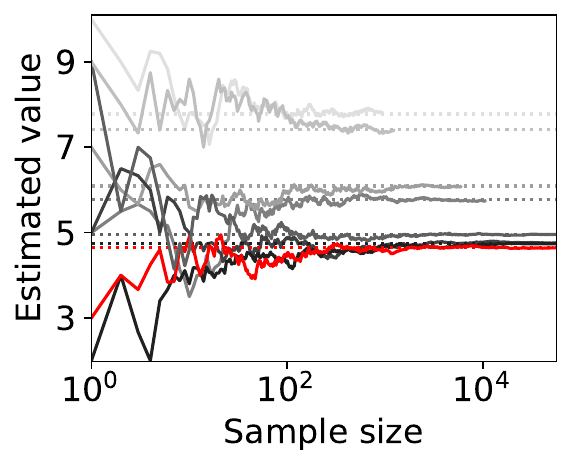}\\
    (a) sample path
    \end{minipage}
    \begin{minipage}{0.33\linewidth}
    \centering
    \includegraphics[width=\linewidth]{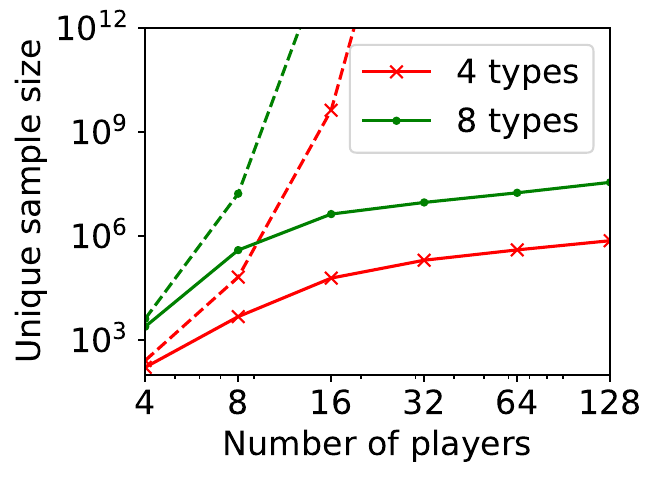}\\
    (b) sample size against $N$
    \end{minipage}
    \begin{minipage}{0.33\linewidth}
    \centering
    \includegraphics[width=\linewidth]{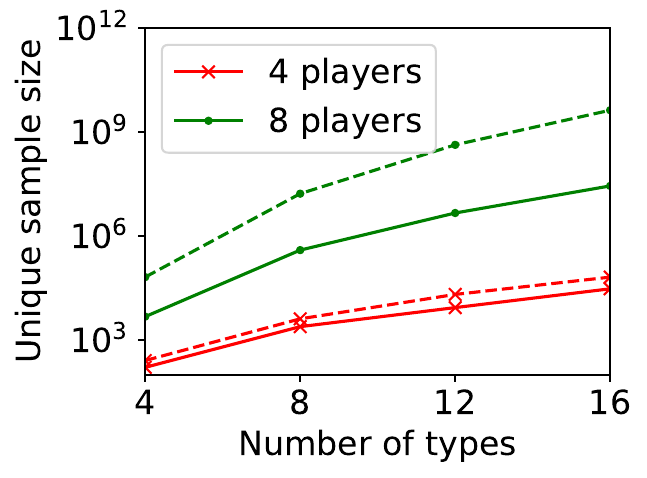}\\
    (c) sample size against $T$
    \end{minipage}
    \caption{
      (a) Representative sample path that shows the estimated values of $\bE[w^\star(t)\mid t_n]$ for $t_n\in\cT_n$ against sample size used by the $(0.25, 0.1)$-PAC SE-BME
      for 8 players, each with 8 types.
      (b)-(c)
      The unique sample size (the number of unique $t$ which $w^\star(t)$ is evaluated with) required by the exact computation of $\min_{t_n\in\cT_n}\bE[w^\star(t)\mid t_n], \forall n\in\cN$ (dashed) and by SE-BME (solid).}
    \label{fig:sample-n-eval}
\end{figure}

For question \romannumeral 1), we assess the effectiveness of SE-BME in reducing the number of evaluations of $w^\star(t)$ within a setting of mechanism design, as detailed in Appendix~\ref{sec:detail:exp:sample}.  Figure~\ref{fig:sample-n-eval} summarizes the results.  Panel (a) shows that SE-BME estimates the means near the minimum value with high accuracy, while quickly eliminating other arms after only a small number of samples, thereby reducing overall sampling.  Panels (b)-(c) demonstrate that SE-BME (solid curves) evaluates $w^\star(t)$ orders of magnitude fewer times than exact computation (dashed curves).  This reduction becomes especially pronounced as the number of players $N$ grows.  While exact computation scales exponentially with $N$, SE-BME scales linearly.  This relative insensitivity of SE-BME to $N$ makes intuitive sense, since increasing $N$ affects the reward distribution but not the number of arms.

\begin{figure}[t]
    \begin{minipage}{0.48\linewidth}
    \centering
    \underline{IR by $\theta \equiv 0$ and \eqref{eq:optimal-eta-IR}-\eqref{eq:solution:rho-theta}.}
    \end{minipage}
    \begin{minipage}{0.02\linewidth}
    \
    \end{minipage}
    \begin{minipage}{0.48\linewidth}
    \centering
    \underline{SBB by $\rho = 0$, \eqref{eq:optimal-eta-IR}-\eqref{eq:optimal-delta-main}, and \eqref{eq:detail:solution:WBB:theta}.}
    \end{minipage}\\
    \begin{minipage}{0.24\linewidth}
    \centering
    \includegraphics[height=\linewidth]{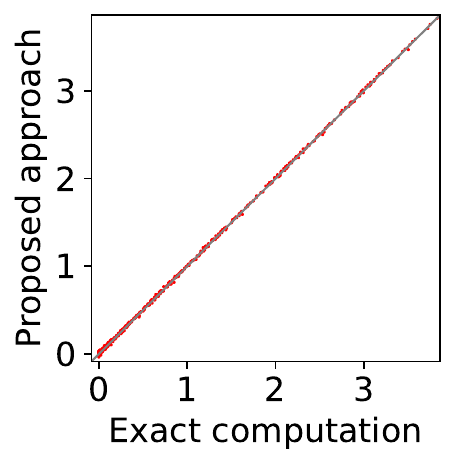}\\
    (a) players' utility
    \end{minipage}
    \begin{minipage}{0.24\linewidth}
    \centering
    \includegraphics[height=\linewidth]{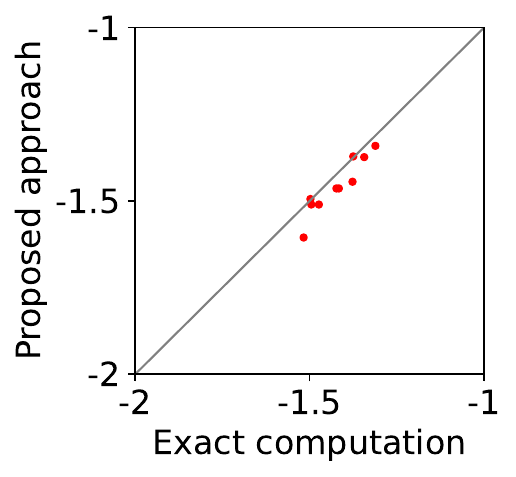}\\
    (b) mediator's revenue
    \end{minipage}
    \begin{minipage}{0.24\linewidth}
    \centering
    \includegraphics[height=\linewidth]{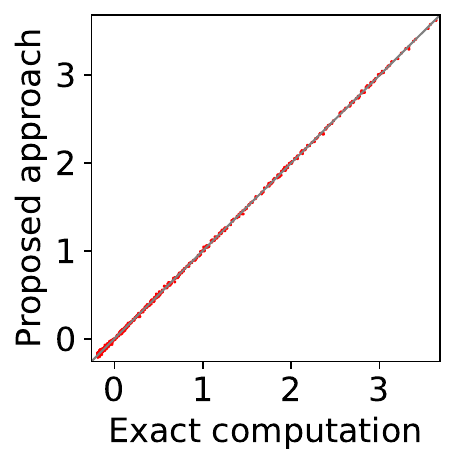}
    (c) players' utility
    \end{minipage}
    \begin{minipage}{0.24\linewidth}
    \centering
    \includegraphics[height=\linewidth]{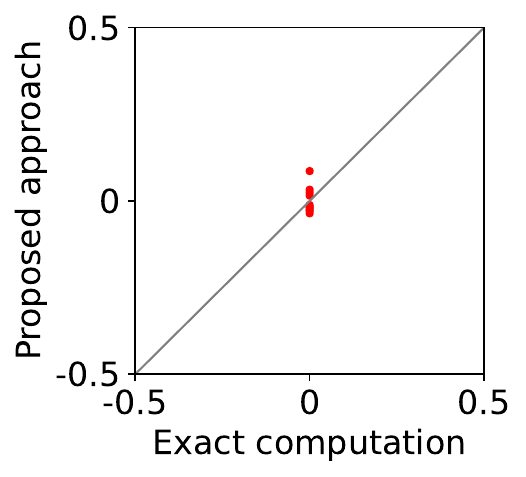}
    (d) mediator's revenue
    \end{minipage}
    \caption{
      The red dot shows expected utility in (a) and (c) and expected revenue in (b) and (d), when analytical solutions are evaluated exactly (horizontal) or estimated with $(0.25,0.1)$-PAC SE-BME (vertical) for environments with $8$ players, each having $8$ possible types.  The analytical solution guarantees IR in (a) and (b) and SBB in (c) and (d).  Results are plotted for 10 random seeds.
    }
    \label{fig:bb-ir-main}
\end{figure}

For question \romannumeral 2), we quantitatively assess how well $\theta$-IR and $\rho$-SBB are preserved when the $\kappa_n(\theta)$ in \eqref{eq:min-heavy} is estimated using SE-BME, where we continue to use the same setting of mechanism design.  Figure~\ref{fig:bb-ir-main} studies two analytical solutions: one that guarantees $\theta$-IR (with $\theta\equiv 0$), and another that guarantees $\rho$-SBB (with $\rho=0$).
In Panels (a) and (c), red dots indicate the expected utility of the players; in Panels (b) and (d), they show the expected revenue of the mediator. Each dot compares exact evaluation (horizontal axis) with SE-BME estimation (vertical axis).
We observe that the mediator’s revenue is more sensitive to estimation error than player utility.
This is because \eqref{eq:LP-WBB} aggregates over all $\eta_n$ terms, whereas \eqref{eq:LP-IR} involves only a single $\eta_n$.
Nevertheless, we can ensure that $\rho$-WBB (rather than SBB) holds with high probability by replacing the $\rho$ with a slightly larger $\rho' > \rho$ when computing $\eta_n$ after estimating $\kappa_n(\theta)$.
As illustrated in Figure~\ref{fig:bb-ir-rho} (Appendix~\ref{sec:detail:exp:ir-bb}), this effectively shifts the dots upward and enables us to guarantee $\rho$-WBB with appropriate tuning of $\rho'$.


\section{Conclusion}
\label{sec:conclusion}

We have analytically derived optimal solutions for the LP that gives mechanisms that guarantee the desired properties of DE, DSIC, $\rho$-SBB, and $\theta$-IR.  When there are $N$ players, each with $T$ possible types, the LP involves $N\,T^{N-1}$ variables, while our solutions are represented by only $N$ essential variables.  While \cite{osogami2023learning} numerically solves this LP only for $N=T=2$, we have exactly evaluated our solutions for $N=T=8$ (see Figure~\ref{fig:sample-n-eval}).  Our solution, however, involves a term whose exact evaluation requires finding efficient social decisions $T^N$ times.  We have modeled the problem of evaluating this term as best mean estimation in multi-armed bandit, proposed a PAC estimator, and proved its asymptotic optimality by establishing a lower bound on the sample complexity.  Our experiments show that our PAC estimator enables finding mechanism for $N=128$ and $T=8$ with a guarantee on the desired properties.


The proposed approach significantly advances the field and offers societal benefits by guaranteeing desired properties in large environments, which existing methods cannot handle. However, it has certain limitations and challenges, which suggest directions for future research. Notably, our mechanisms guarantee $\rho$-SBB and $\theta$-IR only in expectation and have been tested in environments with up to 128 players and 16 types. Additionally, while DE ensures efficiency, it does not guarantee fairness for all players. These limitations and societal impacts are further discussed in Appendix.~\ref{sec:detail:conclusion}.

\section*{Acknowledgements}
We sincerely thank Junya Honda for identifying an error in the proof presented in the previous version.

\bibliographystyle{abbrv}
\bibliography{bandit,game}

\newpage

\newpage
\appendix
\section{Details}
\label{sec:detail}

In this section, we provide full details of derivations and other details skipped in the body of the paper.

\subsection{Details of Section~\ref{sec:LP}}
\label{sec:detail:LP}

Equivalence in \eqref{eq:IRequiv} follows from
\begin{align}
  \lefteqn{
    \bE[ v(\phi^\star(t); t_n) - \tau_n(t) \mid t_n ]
    \ge \theta(t_n)
  }\notag\\
  & \Longleftrightarrow
  \bE\left[ \sum_{m\in\cN} v(\phi^\star(t); t_m) \midd t_n \right]
  - \bE\left[ h_n(t_{-n}) \midd t_n \right]
  \ge \theta(t_n) \\
  & \Longleftrightarrow
  \bE\left[ w^\star(t) \midd t_n \right]
  - \bE\left[ h_n(t_{-n}) \midd t_n \right]
  \ge \theta(t_n).
\end{align}

Equivalence in \eqref{eq:WBBequiv} follows from
\begin{align}
  \sum_{n\in\cN} \bE\left[ \tau_n(t) \right] \ge \rho
  & \Longleftrightarrow
  \sum_{n\in\cN} \bE\left[ h_n(t_{-n})\right]
  - \sum_{n\in\cN} \sum_{m\in\cN_{-n}} \bE\left[ v(\phi^\star(t); t_m) \right] \ge \rho \\
  & \Longleftrightarrow
  \sum_{n\in\cN} \bE\left[ h_n(t_{-n})\right]
  - (N-1) \, \bE\left[ w^\star(t) \right] \ge \rho,
\end{align}
where the last equivalence follows from the definition of $w^\star(t)$ in \eqref{eq:wstar}.

\begin{proposition}
  Let the decision rule $\phi^\star$ be the one that satisfies DE and the payment rule $\tau=(\tau_n)_n$ be in the form of \eqref{eq:VCG-payment} where $h=(h_n)_n=(h_n^\star)_n=h^\star$ is given by the solution to the LP \eqref{eq:LP-obj}-\eqref{eq:LP-WBB}. Then the VCG mechanism $(\phi^\star,\tau^\star)$ satisfies DSIC, DE, $\theta$-IR, and $\rho$-WBB.
  \label{prop:LP}
\end{proposition}
\begin{proof}
  With the equivalences \eqref{eq:IRequiv}-\eqref{eq:WBBequiv}, the constraints \eqref{eq:LP-IR}-\eqref{eq:LP-WBB} in the LP guarantee that $\theta$-IR and $\rho$-WBB are satisfied by feasible solutions. Since we consider the class of VCG mechanisms, DE is trivially satisfied by the definition of $\phi^\star$, and DSIC is satisfied when the payment rule is in the form of \eqref{eq:VCG-payment}. Hence, all of DSIC, DE, $\theta$-IR, and $\rho$-WBB are satisfied by $(\phi^\star,\theta^\star)$.
\end{proof}

Algorithm~\ref{alg:protocol} summarizes the protocol under the VCG mechanism discussed Section~\ref{sec:LP}.  In Step~\ref{step:declare}, the optimal strategy of each player is to truthfully declare its type $\hat t_n=t_n$.  In Step~\ref{step:LP}, the LP may not be feasible, in which case the protocol may fail, or we may use another payment rule to proceed.

\begin{algorithm}[tbh]
  \caption{Protocol under the VCG mechanism}
  \begin{algorithmic}[1]
    \STATE Sample the type profile $t$ from the common prior $\bP$
    \STATE Each player $n\in\cN$ gets to know its own type $t_n$
    \STATE Each player $n\in\cN$ declare their type $\hat t_n$ \label{step:declare}
    \STATE Determine the social decision: $\phi^\star(\hat t) \gets \argmax_{d\in\cD} \displaystyle\sum_{n\in\cN} v(d;\hat t_n)$
    \STATE $h^\star \gets $ Find the optimal solution to the LP \eqref{eq:LP-obj}-\eqref{eq:LP-WBB} \label{step:LP}
    \STATE Determine the payment from each player $n\in\cN$ to the mediator: \\
    $\tau_n(\hat t)
    \gets
    h^\star_n(\hat t_{-n})
    - \displaystyle\sum_{m\in\cN_{-n}}v(\phi^\star(\hat t); \hat t_m)$
    \STATE Each player $n\in\cN$ gets utility
    $v(\phi^\star(\hat t); t_n) - \tau_n(\hat t)$
  \end{algorithmic}
  \label{alg:protocol}
\end{algorithm}

\subsection{Details of Section~\ref{sec:solution}}
\label{sec:detail:solution}

\subsubsection{Solutions that guarantee WBB}
\label{sec:detail:solution:WBB}
Alternatively, one may set
\begin{align}
  \theta(t_n)
  & = \left[\bE[w^\star(t)\mid t_n] - \frac{N-1}{N}\bE[w^\star(t)]\right]^-
  \qquad\forall t_n\in\cT_n, \forall n\in\cN \label{eq:detail:solution:WBB:theta}\\
  \rho & = 0
\end{align}
to guarantee the feasibility, since
\begin{align}
  \lefteqn{
    \sum_{n\in\cN} \min_{t_n\in\cT_n} \left\{\bE[w\mid t_n] - \left[\bE[w^\star(t)\mid t_n] -\frac{N-1}{N}\,\bE[w^\star(t)]\right]^-\right\}
    - (N-1) \, \bE[w^\star(t)]
  }\notag\\
  & =
  \sum_{n\in\cN} \left(
    \min_{t_n\in\cT_n} \left\{\bE[w\mid t_n]-\frac{N-1}{N}\,\bE[w^\star(t)]
    - \left[\bE[w^\star(t)\mid t_n] -\frac{N-1}{N}\,\bE[w^\star(t)]\right]^-\right\}
  \right)\\
  & \ge 0.
\end{align}
In this case, player $n$ may incur negative utility when it has type $t_n$ with $\theta(t_n)<0$, although the loss is guaranteed to be bounded by $|\theta(t_n)|$.

\subsubsection{On dependent types}

The condition \eqref{eq:condition} may be satisfied even when types are dependent, and the optimality of our analytic solutions in Lemma~\ref{lemma:optimal} is guaranteed as long as \eqref{eq:condition} is satisfied.  When types are dependent, however, there are cases where feasible solutions exist even when \eqref{eq:condition} is violated (Lemma~\ref{lemma:condition}).

In the proof of Lemma~\ref{lemma:condition}, we construct such a case with an extreme example of completely dependent types.  However, \eqref{eq:condition} is often satisfied even in such extreme cases of completely dependent types.  For example, as long as
\begin{align}
  x_1 \le x_2 \le \frac{2-p}{1-p}\,x_1,
\end{align}
condition \eqref{eq:condition} is satisfied in the example in the proof of Lemma~\ref{lemma:condition}, since then $(x_1,x_2)$ satisfies \eqref{eq:unneeded-condition}, which corresponds to \eqref{eq:condition} in this example.

\subsection{Details of Section~\ref{sec:bandit2}}
\label{sec:detail:bandit2}

\subsubsection{Upper bound}

\begin{algorithm}[t]
  \caption{Best Mean Estimator}
  \begin{algorithmic}[1]
    \STATE $\hat I \gets \texttt{PAC-BAI}(\varepsilon,\delta,T)$ \label{step:best-reward:bai}
    \STATE Pull arm $\hat I$ for $m^\star$ times \label{step:best-reward:pull}
    \STATE Let $\hat\mu_{\hat I}$ be the average of the $m^\star$ samples
    \RETURN $\hat\mu_{\hat I}$
  \end{algorithmic}
  \label{alg:best-reward}
\end{algorithm}

We consider Algorithm~\ref{alg:best-reward} that uses an arbitrary $(\varepsilon,\delta)$-PAC BAI to identify the best arm $\hat I$ and take a predetermined number $m^\star$ of samples from $\hat I$ to estimate its mean.
Notice that the naive approach of sampling each arm $\Theta((1/\varepsilon^2)\,\log(1/\delta))$ times, which also trivially falls within the upper bound in Theorem~\ref{thrm:complexity-bme}, would only guarantee that the best mean is estimated with the error bound of $\varepsilon$ with probability at least $(1-\delta)^K$.
Conversely, it would require sampling each arm $\Omega((1/\varepsilon^2)\,\log(K/\delta))$ times to obtain the same error bound with probability $1-\delta$.

\subsubsection{Lower bound}

We derive the lower bound in Lemma~\ref{lemma:lower} by reducing BME to the $\varepsilon$-Biased Coin Problem:
\begin{definition}[$\varepsilon$-Biased Coin Problem]
  For $0<\varepsilon<1$, consider a Bernoulli random variable $B$ whose mean $\alpha$ is known to be either $\alpha^+\coloneqq(1+\varepsilon)/2$ or $\alpha^-\coloneqq(1-\varepsilon)/2$.  The $\varepsilon$-Biased Coin Problem asks to correctly identify whether $\alpha=\alpha^+$ or $\alpha=\alpha^-$.
  \label{def:bias}
\end{definition}

A lower bound on the sample complexity for solving the $\varepsilon$-Biased Coin Problem is known as the following lemma, for which we provide a proof in Appendix~\ref{sec:proof} for completeness:
\begin{lemma}[\cite{chernoff1972sequential,evendar02pac}]
  For $0<\delta<1/2$, any algorithm that solves the $\varepsilon$-Biased Coin Problem correctly with probability at least $1-\delta$ has expected sample complexity at least $\Omega((1/\varepsilon^2)\,\log(1/\delta))$.
  \label{lemma:bias}
\end{lemma}

We have derived the lower bound for BME
using the technique used for a lower bound for BAI in \cite{evendar02pac}; see the proof of Lemma~\ref{lemma:lower} in Appendix~\ref{sec:proof:bandit2}. However, as we have discussed at the end of Section~\ref{sec:bandit2}, while our lower bound for BME is tight, the lower bound for BAI in \cite{evendar02pac} is not.  The difference in the derived lower bound stems from the following behavior of BME and BAI when all the arms have mean reward of $\alpha^-$ and hence are indistinguishable.  The algorithm constructed in \cite{evendar02pac} determines that $B$ has mean $\alpha^+$ when either arm $i^+$ or arm $i^-$ is identified as the best arm.  When the arms are indistinguishable, a PAC BAI would correctly identify each of the $K$ arms, including $i^+$ or $i^-$, as the best arm uniformly at random, which induces an error with probability $1/K$.  On the other hand, the mean reward estimated by a PAC BME would be approximately correct with high probability, even when the arms are indistinguishable.

\subsection{Details of Section~\ref{sec:exp}}
\label{sec:detail:exp}

In this section, we conduct several numerical experiments to validate the effectiveness of the proposed approach and to understand its limitations.  Specifically, we design our experiments to investigate the following three questions:
\begin{enumerate}
  \item BME studied in Section~\ref{sec:bandit} has asymptotically optimal sample complexity, but how well can we estimate the best mean when there are only a moderate number of arms?
  \item How much can BME reduce the number of times $w^\star(t)$ given by \eqref{eq:wstar} is evaluated?
  \item How well $\theta$-IR and $\rho$-SBB are satisfied when \eqref{eq:min-heavy} is estimated by BME rather than calculated exactly?
\end{enumerate}

\subsubsection{Best Mean Estimation with moderate number of arms}
\label{sec:detail:exp:moderate}

For BAI, existing algorithms that have asymptotically optimal sample complexity often perform poorly with moderate number of arms \cite{hassidim2020optimal}.  As a result, Approximate Best Arm \cite{hassidim2020optimal}, which has optimal sample complexity, runs Naive Elimination, which has asymptotically suboptimal sample complexity, when the number of arms is below $10^5$ or after sufficient number of suboptimal arms is eliminated.  The BME algorithm studied in Section~\ref{sec:bandit} has asymptotically optimal sample complexity but relies on BAI, and thus its sample complexity for moderate number of arms is not well characterized by Theorem~\ref{thrm:complexity-bme}.  Similar to BAI, BME relies on an algorithm that performs well when there are only moderate number of arms.

\begin{algorithm}[t]
  \caption{Successive Elimination for Best Mean Estimation ($(\varepsilon,\delta)$-PAC SE-BME)}
  \begin{algorithmic}[1]
    \REQUIRE $\varepsilon,\delta$
    \STATE Let $\cR\gets\{1,\ldots,K\}$ be the set of remaining arms
    \STATE Let $t\gets 0; \alpha\gets 1$
    \WHILE{$\alpha>\varepsilon$}
    \STATE Pull each arm $k\in \cR$ once and update the sample average $\hat\mu_k$
    \STATE $t \gets t+1$
    \STATE $\alpha\gets\sqrt{\frac{1}{2\,t} \log\left(\frac{\pi^2\,K\,t^2}{3\,\delta}\right)}$
    \FORALL{$k\in \cR$}
    \STATE Remove $k$ from $\cR$ if $\max_{\ell\in \cR} \hat\mu_\ell - \hat\mu_k \ge 2 \, \alpha$
    \ENDFOR
    \ENDWHILE
    \RETURN $\max_{k\in \cR} \hat\mu_k$
  \end{algorithmic}
  \label{alg:SE-BME}
\end{algorithm}

\begin{algorithm}[t]
  \caption{Successive Elimination for Best Arm Identification ($(\varepsilon,\delta)$-PAC SE-BAI)}
  \begin{algorithmic}[1]
    \REQUIRE $\varepsilon,\delta$
    \STATE Let $\cR\gets\{1,\ldots,K\}$ be the set of remaining arms
    \STATE Let $t\gets 0; \alpha\gets 1$
    \WHILE{$|\cR|>1$ and $\alpha>\frac{\varepsilon}{2}$}
    \STATE Pull each arm $k\in \cR$ once and update the sample average $\hat\mu_k$
    \STATE $t \gets t+1$
    \STATE $\alpha\gets\sqrt{\frac{1}{2\,t}\log\left(\frac{\pi^2\,K\,t^2}{6\,\delta}\right)}$
    \FORALL{$k\in \cR$}
    \STATE Remove $k$ from $\cR$ if $\max_{\ell\in \cR} \hat\mu_\ell - \hat\mu_k \ge 2 \, \alpha$
    \ENDFOR
    \ENDWHILE
    \RETURN $\argmax_{k\in \cR} \hat\mu_k$
  \end{algorithmic}
  \label{alg:SE-BAI}
\end{algorithm}

In this section, we compare the performance of Successive Elimination algorithms for BAI and BME.  Successive Elimination has suboptimal sample complexity, similar to Naive Elimination, but often outperforms Naive Elimination for moderate number of arms \cite{evendar06action}.  Specifically, we compare the performance of $(\varepsilon,\delta)$-PAC Successive Elimination for BME (SE-BME; Algorithm~\ref{alg:SE-BME}) against the corresponding $(\varepsilon,\delta)$-PAC Successive Elimination for BAI (SE-BAI; Algorithm~\ref{alg:SE-BAI}).  For completeness, in Appendix~\ref{sec:proof}, we provide standard proofs on the correctness of these algorithms, as stated in Proposition~\ref{prop:se-bme} and Proposition~\ref{prop:se-bai}:

\begin{proposition}
  Algorithm~\ref{alg:SE-BME} is an $(\varepsilon,\delta)$-PAC BME.
  \label{prop:se-bme}
\end{proposition}

\begin{proposition}
  Algorithm~\ref{alg:SE-BAI} is an $(\varepsilon,\delta)$-PAC BAI.
  \label{prop:se-bai}
\end{proposition}

There are three differences between SE-BME and SE-BAI.  First, SE-BME exits the while-loop when $\alpha<\varepsilon$ (while SE-BAI needs to wait until $\alpha<\varepsilon/2$), since it can return an overestimated mean of a suboptimal arm as long as the estimated value is within $\varepsilon$ from the best mean.  Second, SE-BAI exits the while-loop when only one arm remains, since it does not need to precisely estimate the mean of the identified arm.  Finally, SE-BAI uses smaller $\alpha_t$, since it only needs to consider one-sided estimation error (underestimation for best arms and overestimation for other arms).

\begin{figure}[t]
  \centering
  \begin{minipage}{0.1\linewidth}
    $\delta=0.05$
  \end{minipage}
  \begin{minipage}{0.29\linewidth}
    \includegraphics[width=\linewidth]{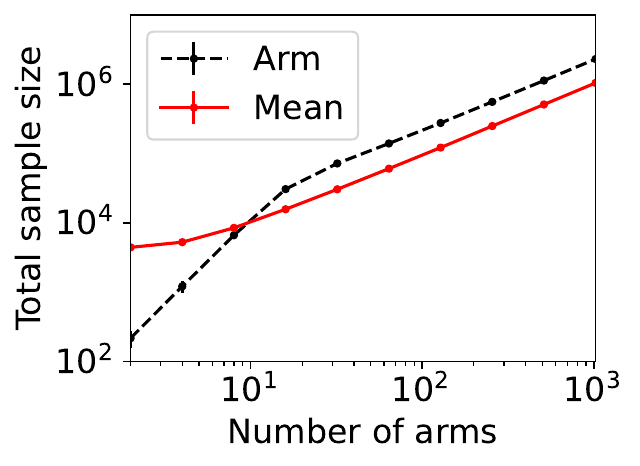}
  \end{minipage}
  \begin{minipage}{0.29\linewidth}
    \includegraphics[width=\linewidth]{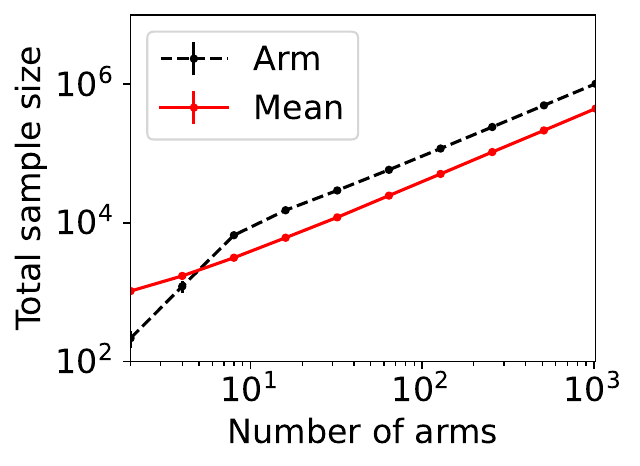}
  \end{minipage}
  \begin{minipage}{0.29\linewidth}
    \includegraphics[width=\linewidth]{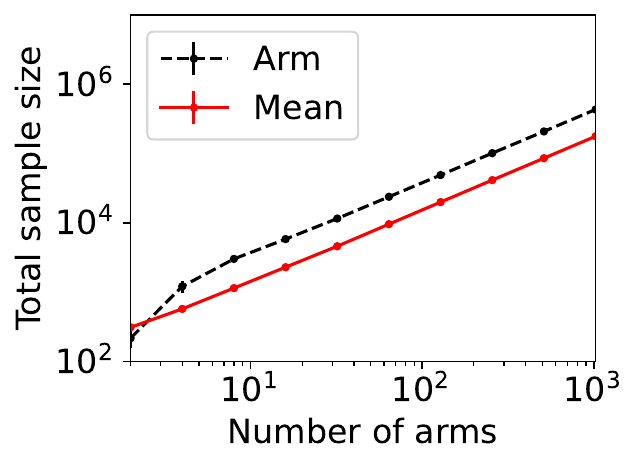}
  \end{minipage}
  \begin{minipage}{0.1\linewidth}
    $\delta=0.1$
  \end{minipage}
  \begin{minipage}{0.29\linewidth}
    \centering
    \includegraphics[width=\linewidth]{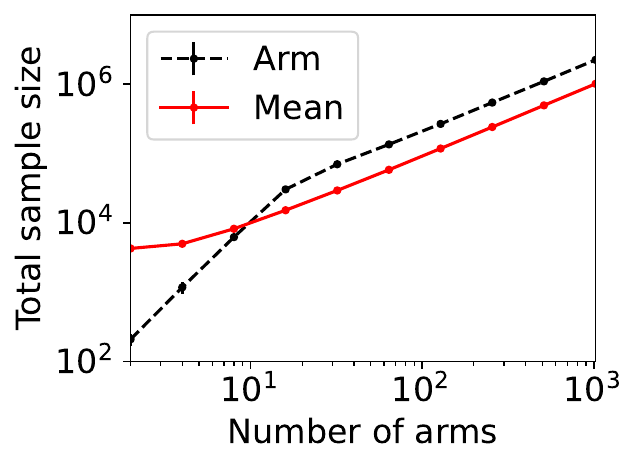}
  \end{minipage}
  \begin{minipage}{0.29\linewidth}
    \centering
    \includegraphics[width=\linewidth]{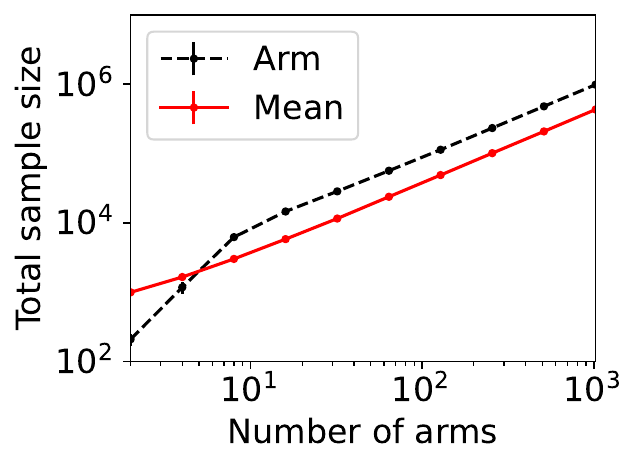}
  \end{minipage}
  \begin{minipage}{0.29\linewidth}
    \centering
    \includegraphics[width=\linewidth]{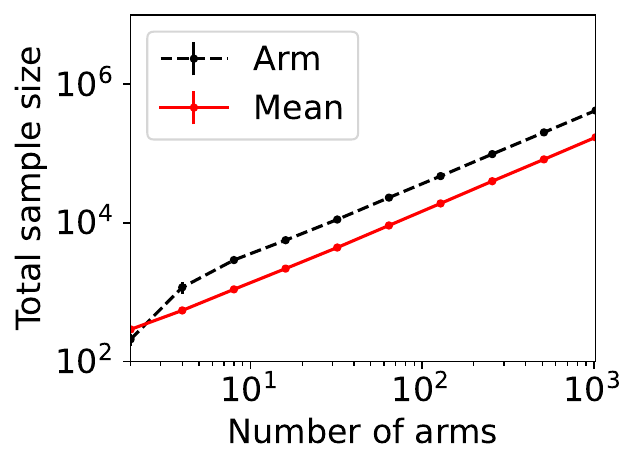}
  \end{minipage}
  \begin{minipage}{0.1\linewidth}
    $\delta=0.2$
  \end{minipage}
  \begin{minipage}{0.29\linewidth}
    \centering
    \includegraphics[width=\linewidth]{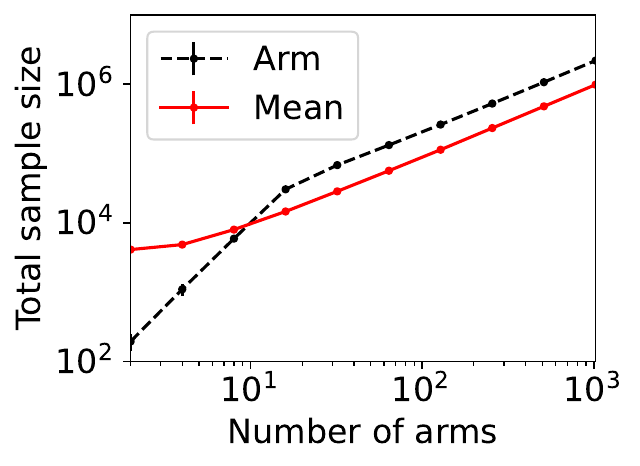}
    (a) $\varepsilon=0.05$
  \end{minipage}
  \begin{minipage}{0.29\linewidth}
    \centering
    \includegraphics[width=\linewidth]{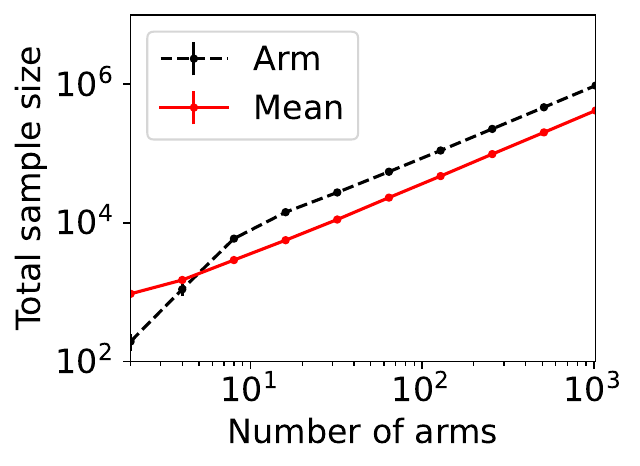}
    (b) $\varepsilon=0.1$
  \end{minipage}
  \begin{minipage}{0.29\linewidth}
    \centering
    \includegraphics[width=\linewidth]{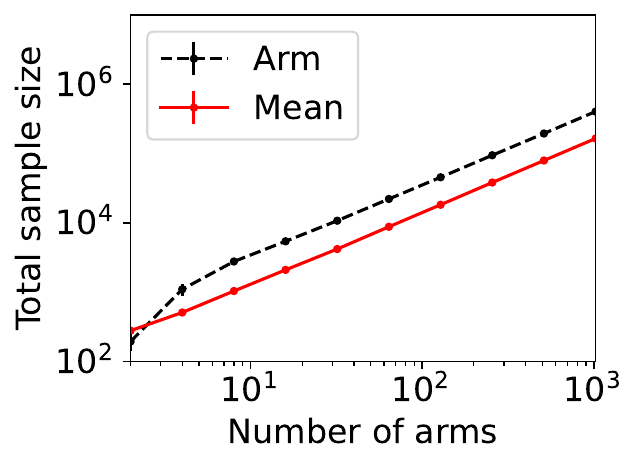}
    (c) $\varepsilon=0.2$
  \end{minipage}
  \caption{The total sample size required by $(\varepsilon,\delta)$-PAC BME (Mean; Algorithm~\ref{alg:SE-BME}) and by $(\varepsilon,\delta)$-PAC BAI (Arm; Algorithm~\ref{alg:SE-BAI}) when arms have Bernoulli rewards with equally separated means for varying values of $\varepsilon$ and $\delta$.}
  \label{fig:BME-BAI}
\end{figure}

Figure~\ref{fig:BME-BAI} shows the total sample size required by SE-BME and by SE-BAI for varying values of $\varepsilon$ and $\delta$, and for varying number $K$ of arms.  Here, the arms have Bernoulli rewards, and their means are selected in a way they are equally separated (i.e., $\mu_k=(k-0.5)/K$ for $k=1,\ldots,K$).  For each datapoint, the experiments are repeated 10 times.  The standard deviation of the total sample size is plotted but too small to be visible in the figure.

Overall, it can be observed that SE-BME generally requires smaller sample size than SE-BAI, except when there are only a few arms (and the means have large gaps in the setting under consideration).  The efficiency of SE-BAI for a small number of arms makes intuitive sense, because the best arm can be identified without estimating the means with high accuracy.

\subsubsection{Effectiveness of Best Mean Estimation in mechanism design}
\label{sec:detail:exp:sample}

In this section, we quantitatively validate the effectiveness of BME in reducing the number of times $w^\star(t)$ is computed when we evaluate \eqref{eq:min-heavy}.  We use $(\varepsilon,\delta)$-PAC SE-BME (Algorithm~\ref{alg:SE-BME}) as BME.

To this end, we consider the following mechanism design, motivated by the double-sided auctions for electricity \cite{zou2009double,hobbs2000evaluation}, where we have $N=|\cN|$ players, each player $n\in\cN$ has $K$ possible types (i.e., $|\cT_n|=K$), with varying values of $N$ and $K$.  The $K$ possible types of each player are selected uniformly at random from integers, $[-K, K]$, without replacement.  We then assume, as the common prior $\bP$, that the type of each player is distributed uniformly among the $K$ possible types and independent of the types of other players.  Each player is a buyer or a seller of a single item, depending on its type.  When a player $n$ has a positive type $t_n$, the player is a buyer who wants to buy a unit, whose valuation to the player is $t_n$ (i.e., $v(d;t_n)=t_n$ if player $n$ buys a unit of the item with social decision $d$).  When the player has a negative type $t_n$, the player is a seller who wants to sell a unit, which incurs cost $|t_n|$ (i.e., $v(d;t_n)=t_n$ if player $n$ sells a unit of the item with social decision $d$).  The player does not participate in the market, when its type is zero.  For a given profile of types $t$, a social decision is given by a bipartite matching between buyers and sellers.  In this setting, we can immediately obtain the social decision with DE by greedily matching buyers of high valuations to sellers of low costs as long as the value of the buyers are higher than the costs of the sellers.

\begin{figure}[t]
  \centering
  \begin{minipage}{0.11\linewidth}
    4 players
  \end{minipage}
  \begin{minipage}{0.29\linewidth}
    \includegraphics[width=\linewidth]{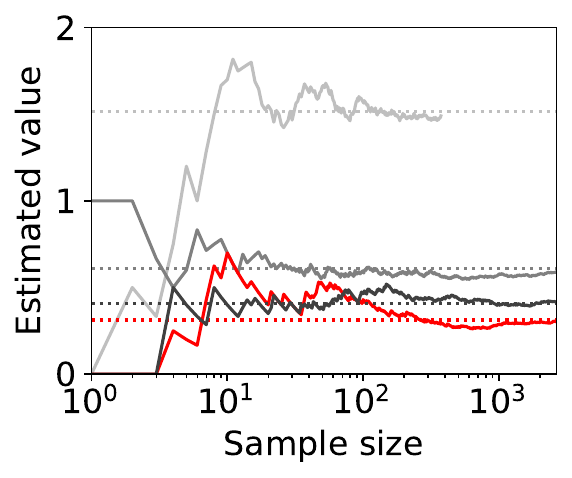}
  \end{minipage}
  \begin{minipage}{0.29\linewidth}
    \includegraphics[width=\linewidth]{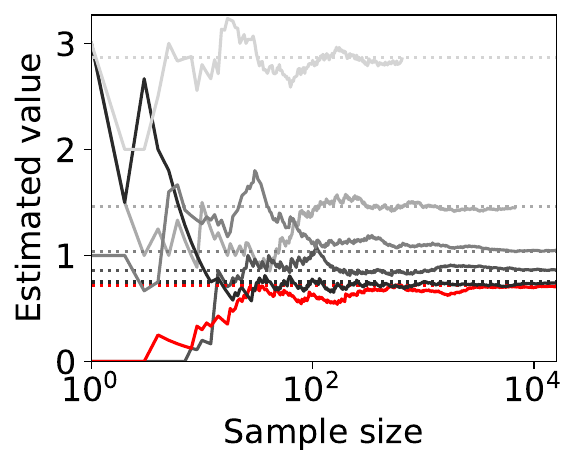}
  \end{minipage}
  \begin{minipage}{0.29\linewidth}
    \includegraphics[width=\linewidth]{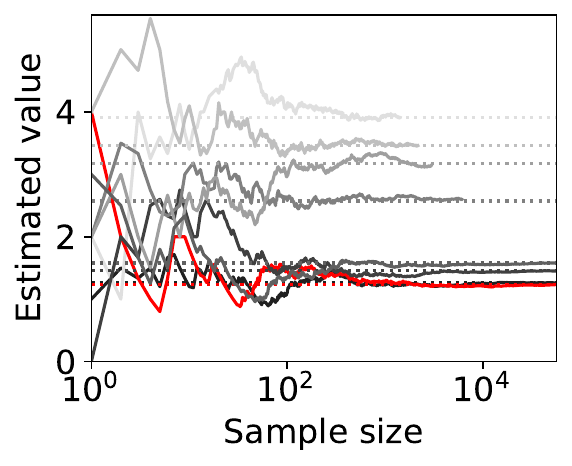}
  \end{minipage}
  \begin{minipage}{0.11\linewidth}
    6 players
  \end{minipage}
  \begin{minipage}{0.29\linewidth}
    \centering
    \includegraphics[width=\linewidth]{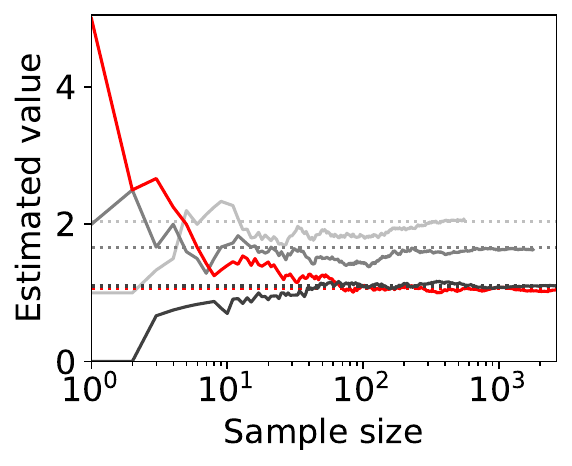}
  \end{minipage}
  \begin{minipage}{0.29\linewidth}
    \centering
    \includegraphics[width=\linewidth]{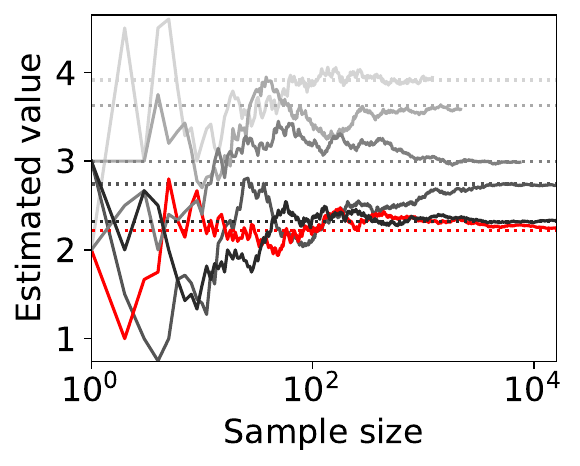}
  \end{minipage}
  \begin{minipage}{0.29\linewidth}
    \centering
    \includegraphics[width=\linewidth]{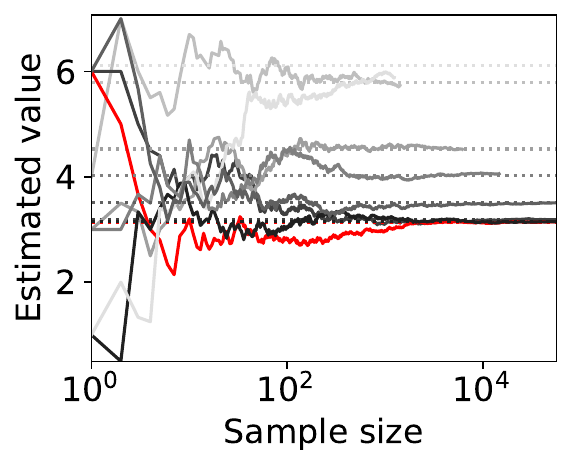}
  \end{minipage}
  \begin{minipage}{0.11\linewidth}
    8 players
  \end{minipage}
  \begin{minipage}{0.29\linewidth}
    \centering
    \includegraphics[width=\linewidth]{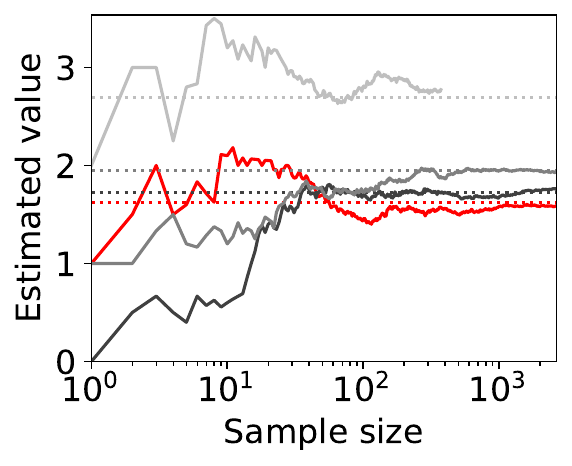}
    (a) 4 types
  \end{minipage}
  \begin{minipage}{0.29\linewidth}
    \centering
    \includegraphics[width=\linewidth]{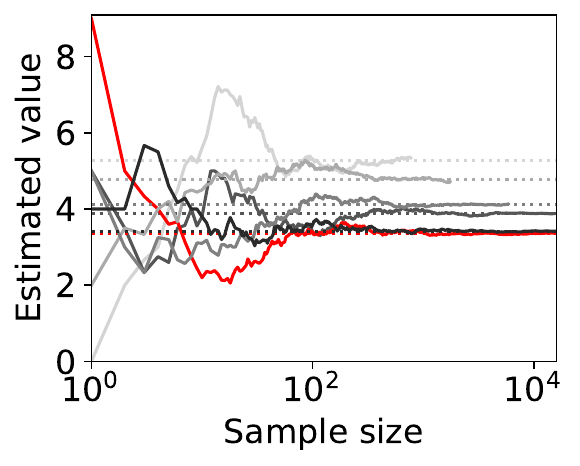}
    (b) 6 types
  \end{minipage}
  \begin{minipage}{0.29\linewidth}
    \centering
    \includegraphics[width=\linewidth]{figs/bandit_8_8_7_0.5_0.1.pdf}
    (c) 8 types
  \end{minipage}
  \caption{Representative sample paths that show the estimated values of $\bE[w^\star(t)\mid t_n]$ for $t_n\in\cT_n$ against sample size used by $(0.25,0.1)$-PAC SE-BME.}
  \label{fig:sample}
\end{figure}

Figure~\ref{fig:sample} shows representative sample paths of the estimated values of $\bE[w^\star(t)\mid t_n]$ for each $t_n\in\cT_n$ when $(0.25,0.1)$-PAC SE-BME\footnote{Minimization in Algorithm~\ref{alg:SE-BME} is translated into maximization.} is used to evaluate $\min_{t_n\in\cT_n} \bE[w^\star(t)\mid t_n]$ (i.e., \eqref{eq:min-heavy} with $\theta(t_n)=0$).  Each panel in Figure~\ref{fig:sample} shows $|\cT_n|$ curves, where the red curve corresponds to the one with minimum $\bE[w^\star(t)\mid t_n]$.

Observe that the types $t_n$ (arms) that have close to the minimum mean, $\min_{t_n\in\cT_n} \bE[w^\star(t)\mid t_n]$, survive until SE-BME terminates, and their means are evaluated with sufficient accuracy.  On the other hand, the types that have large means are eliminated after a relatively small number of samples without being estimated precisely, which contributes to reducing the number of evaluating the value of the efficient social decision $w^\star(t)$.

\begin{figure}[t]
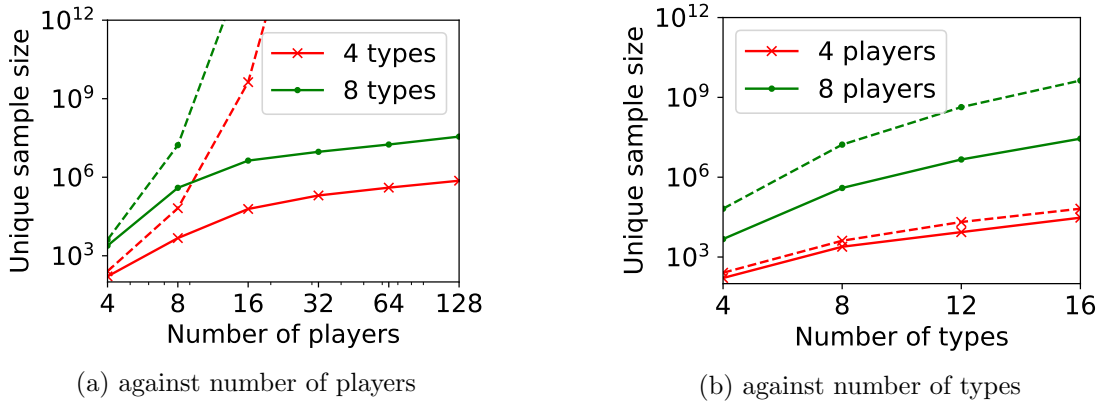

  \centering
  \begin{minipage}{0.4\linewidth}
    \centering
    \includegraphics[width=\linewidth]{figs/bandit_trend_player_128_8_0.25_0.1.pdf}
    (a) against number of players
  \end{minipage}
  \begin{minipage}{0.08\linewidth}
    \
  \end{minipage}
  \begin{minipage}{0.4\linewidth}
    \centering
    \includegraphics[width=\linewidth]{figs/bandit_trend_type_8_16_0.25_0.1.pdf}
    (b) against number of types
  \end{minipage}
  \caption{The unique sample size (the number of unique $t$ which $w^\star(t)$ is evaluated with) required by the exact computation of $\min_{t_n\in\cT_n}\bE[w^\star(t)\mid t_n], \forall n\in\cN$ (dashed curves) and by $(0.25,0.1)$-PAC SE BME (solid curves).}
  \label{fig:n-eval}
\end{figure}

In Figure~\ref{fig:n-eval}, we study how much SE-BME can reduce the number of evaluations of $w^\star(t)$ needed to estimate $\min_{t_n\in\cT_n}\bE[w^\star(t)\mid t]$ for all $n\in\cN$ with the accuracy that is shown in Figure~\ref{fig:sample} (i.e., using the same values of $\varepsilon=0.25$ and $\delta=0.1$).  Recall that $K=|\cT_n|, \forall n\in\cN$ in the setting under consideration.  Hence, the exact computation of $\min_{t_n\in\cT_n}\bE[w^\star(t)\mid t_n]$ for a single $n\in\cN$ would require evaluating $w^\star(t)$ for $K^N$ different values of $t$, but $K^N$ evaluations of $w^\star(t)$ are also sufficient to exactly compute $\min_{t_n\in\cT_n}\bE[w^\star(t)\mid t_n]$ for all $n\in\cN$, because we can cache the value of $w^\star(t)$ and reuse it when it is needed.  Since the computational complexity associated with evaluating $w^\star(t)$ with \eqref{eq:wstar} is the bottleneck, the unique sample size is what we should be interested in.  Similar to exact computation, SE-BME also benefits from caching and reusing the values of $w^\star(t)$.  Figure~\ref{fig:n-eval} compares the unique sample size required by the exact computation and $(0.25,0.1)$-PAC SE-BME.

Figure~\ref{fig:n-eval}(a) implies that SE-BME (shown with solid curves) evaluates the total value of the efficient social decision, $w^\star(t)$, by orders of magnitude smaller number of times than what is required by exact computation (shown with dashed curves).  While the number of evaluations of $w^\star(t)$ grows exponentially with the number of players (specifically, $K^N$) when exact computation is used, it grows only polynomially (in fact, slightly slower than linearly) when SE-BME is used.  This relative insensitivity of the sample complexity of SE-BME to the number of players makes intuitive sense, because the number of players only affects the distribution of the reward and keeps the number of arms unchanged.

Figure~\ref{fig:n-eval}(b) shows the number of evaluations of $w^\star(t)$ against the number of types $K=|\cT_n|$ for any $n\in\cN$.  The advantage of SE-BME over exact computation is relatively minor when we increase the number of types instead of the number of players, since increasing the number of types directly increases the number of arms.  In all cases, however, we can observe that SE-BME can significantly reduce the unique sample size.

\begin{figure}[t]
  \centering
  \begin{minipage}{0.4\linewidth}
    \centering
    \includegraphics[width=\linewidth]{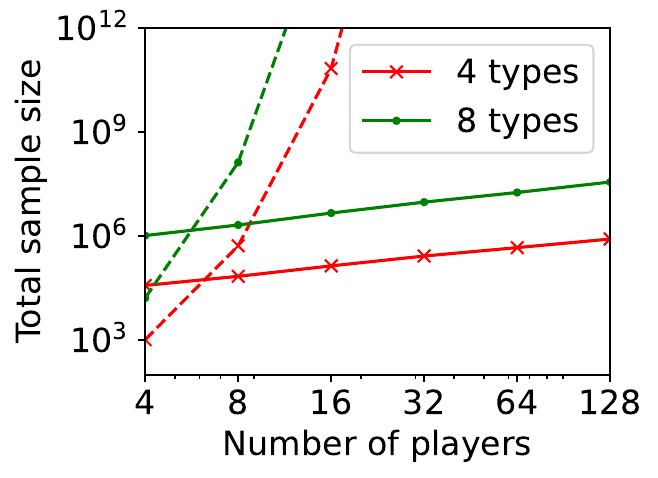}
    (a) against number of players
  \end{minipage}
  \begin{minipage}{0.08\linewidth}
    \
  \end{minipage}
  \begin{minipage}{0.4\linewidth}
    \centering
    \includegraphics[width=\linewidth]{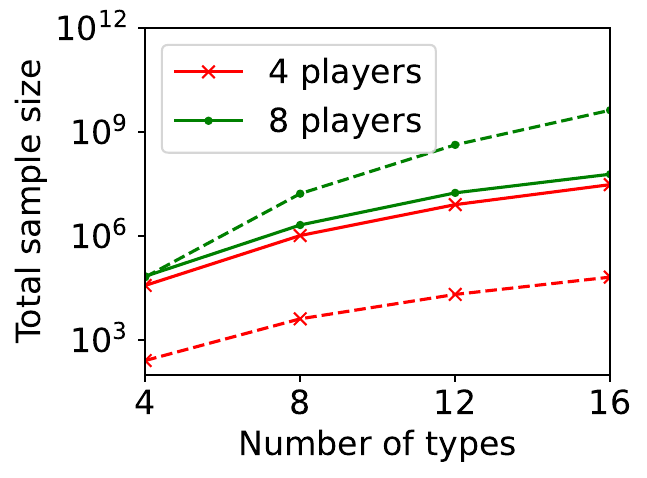}
    (b) against number of types
  \end{minipage}
  \caption{The total sample size (the number of $t$ which $w^\star(t)$ is evaluated with) required by the exact computation of $\min_{t_n\in\cT_n}\bE[w^\star(t)\mid t_n], \forall n\in\cN$ (dashed curves) and by $(0.25,0.1)$-PAC SE-BME (solid curves).}
  \label{fig:n-eval-total}
\end{figure}

Figure~\ref{fig:n-eval-total} shows the total sample size, rather than the unique sample size, required by exact computation (dashed curves) and SE-BME (solid curves).  The total sample size with exact computation is $N\,K^N$.  While $w^\star(t)$ is evaluated $N$ times for each $t$ with exact computation, SE-BME may waste evaluating the same $w^\star(t)$ more often particularly when there are only a small number of players.  This reduces benefits of SE-BME for total sample size, as compared to the unique sample size.

\subsubsection{Individual rationality and budget balance with Best Mean Estimation}
\label{sec:detail:exp:ir-bb}

We next address the question of how well $\theta$-IR and $\rho$-SBB are guaranteed when we estimate $\min_{t_n\in\cT_n} \{ \bE[w^\star(t)\mid t_n] - \theta(t_n) \}$ with BME rather than computing it exactly.  We continue to use the setting of mechanism design introduced in Section~\ref{sec:detail:exp:sample}.  Recall that $\theta$-IR is guaranteed when $\eta_n$ is given by \eqref{eq:optimal-h} and \eqref{eq:optimal-eta-IR}, and $\rho$-SBB is guaranteed when $\eta_n$ is given by \eqref{eq:optimal-h} and \eqref{eq:optimal-delta}.  However, these are guaranteed only when the expected values, $\bE[w^\star(t)]$ and $\bE[w^\star(t)\mid t_n]$, are exactly computed.  In this section, we quantitatively evaluate how well $\theta$-IR and $\rho$-SBB  are satisfied when those expected values are estimated from samples.  Throughout this section, we study the case with $\rho=0$ and $\theta\equiv 0$ (i.e., $\theta(t_n)=0, \forall t_n\in\cT_n, \forall n\in\cN$).

\begin{figure}[t]
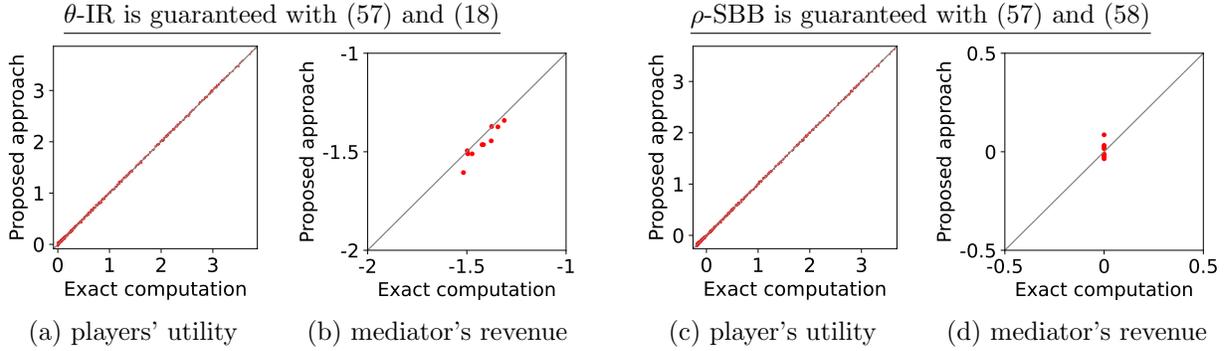

  \begin{minipage}{0.48\linewidth}
    \centering
    \underline{IR by $\theta \equiv 0$ and \eqref{eq:optimal-eta-IR}-\eqref{eq:solution:rho-theta}.}
  \end{minipage}
  \begin{minipage}{0.02\linewidth}
    \
  \end{minipage}
  \begin{minipage}{0.48\linewidth}
    \centering
    \underline{SBB by $\rho = 0$, \eqref{eq:optimal-eta-IR}-\eqref{eq:optimal-delta-main}, and \eqref{eq:detail:solution:WBB:theta}.}
  \end{minipage}\\
  \begin{minipage}{0.24\linewidth}
    \centering
    \includegraphics[height=0.91\linewidth]{figs/amd_ir_8_8_0.25_0.1_0_IRok.pdf}
    (a) players' utility
  \end{minipage}
  \begin{minipage}{0.24\linewidth}
    \centering
    \includegraphics[height=0.91\linewidth]{figs/amd_bb_8_8_0.25_0.1_0_IRok.pdf}
    (b) mediator's revenue
  \end{minipage}
  \begin{minipage}{0.02\linewidth}
    \
  \end{minipage}
  \begin{minipage}{0.24\linewidth}
    \centering
    \includegraphics[height=0.91\linewidth]{figs/amd_ir_8_8_0.25_0.1_0_BBok.pdf}
    (c) player's utility
  \end{minipage}
  \begin{minipage}{0.24\linewidth}
    \centering
    \includegraphics[height=0.91\linewidth]{figs/amd_bb_8_8_0.25_0.1_0_BBok.pdf}
    (d) mediator's revenue
  \end{minipage}
  \caption{The red dots show the expected utility of the players in Columns (a) and (c) and the expected revenue of the mediator in Columns (b) and (d), where analytical solutions are evaluated exactly (horizontal axes) or estimated with $(0.25,0.1)$-PAC SE-BME (vertical axes) for environments with $|\cN|=8$ players, each having $|\cT_n|=8$ possible types.  The analytical solution guarantees $\theta$-IR with $\theta\equiv 0$ in Columns (a) and (b) and $\rho$-SBB with $\rho=0$ in Columns (c) and (d).  Results are plotted for 10 random seeds.  Diagonal lines are also plotted to help understand where the horizontal and vertical axes are equal.}
  \label{fig:bb-ir}
\end{figure}

In Figure~\ref{fig:bb-ir}, we first evaluate the best mean, $\min_{t_n\in\cT_n} \bE[w^\star(t)\mid t]$, either with exact computation or with BME, then compute $\eta_n$ with \eqref{eq:optimal-h} and \eqref{eq:optimal-eta-IR} for Columns (a)-(b) and with \eqref{eq:optimal-h} and \eqref{eq:optimal-delta} for Columns (c)-(d), and finally evaluate the expected utility of each player (the left-hand side of \eqref{eq:LP-IR}) for Columns (a) and (c) and the expected revenue of the mediator (the left-hand side of \eqref{eq:LP-WBB}) for Columns (b) and (d) by setting $h_n(t_{-n})=\eta_n, \forall t_{-n}\in\cT_{-n}, \forall n\in\cN$.  Even if the best mean is estimated with BME, we evaluate the expectations on the left-hand side of \eqref{eq:LP-IR} and \eqref{eq:LP-WBB} with exact computation, because these are the expected utility and the expected revenue that the players and the mediator will experience.  Here, we repeat the experiment with 10 different random seeds, so that there are 10 data-points in Columns (b) and (d), and each of Columns (a) and (c) has $10\times 8 \times 8 = 640$ data-points, where each data-point corresponds to a player of a particular type with a particular random seed.

Overall, Columns (a) and (c) of Figure~\ref{fig:bb-ir} show that the expected utility experienced by the players is relatively insensitive to whether the best mean is evaluated with exact computation (horizontal axes) or with BME (vertical axes).  Taking a closer look, we can observe that, in this particular setting, 0-IR is violated for some players in Column (c) even if the best mean is computed exactly, while it is guaranteed for any player of any type in Column (a) if the best mean is computed exactly.

On the other hand, as is shown in Columns (b) and (d), the mediator experiences non-negligible difference in its expected revenue depending on whether the best mean is evaluated with exact computation or with BME.  It is to be expected that the mediator experiences relatively larger variance in its expected revenue, because \eqref{eq:LP-WBB} involves the summation $\sum_{n\in\cN} \eta_n$, while \eqref{eq:LP-IR} only involves $\eta_n$ for a single $n\in\cN$. In particular, it is evident in Column (d) that 0-WBB (let alone 0-SBB) is violated when the best mean is estimated with BME (vertical axes), while it is always guaranteed with exact computation (horizontal axes).  In Column (b), 0-WBB is violated regardless of whether the best mean is computed exactly or with BME, since satisfying 0-WBB and 0-IR for all players (together with DE and DSIC) is impossible in this particular setting.

A simple remedy to this violation of $\rho$-WBB is to replace the $\rho$ with a $\rho'>\rho$ when we compute the $\eta_n$ from the best means, $\min_{t_n\in\cT_n} \bE[w^\star(t)\mid t_n]$ for $n\in\cN$, estimated with BME.  By considering the ($\varepsilon, \delta$)-PAC guarantee for the error in the estimation, we can guarantee that $\rho$-WBB is satisfied with high probability by setting an appropriate value of $\rho'$.  Analogously, we may replace the $\theta(t_n)$ with a $\theta'(t_n)>\theta(t_n)$ to provide a guaranteed that $\theta$-IR is satisfied with high probability when the best mean is estimated with BME.  See Theorem~\ref{thrm:connect}.

\begin{figure}[t]
  \begin{minipage}{0.48\linewidth}
    \centering
    \underline{IR by $\theta \equiv 0$ and \eqref{eq:optimal-eta-IR}-\eqref{eq:solution:rho-theta}.}
  \end{minipage}
  \begin{minipage}{0.02\linewidth}
    \
  \end{minipage}
  \begin{minipage}{0.48\linewidth}
    \centering
    \underline{SBB by $\rho = 0$, \eqref{eq:optimal-eta-IR}-\eqref{eq:optimal-delta-main}, and \eqref{eq:detail:solution:WBB:theta}.}
  \end{minipage}\\
  \begin{minipage}{0.24\linewidth}
    \centering
    \includegraphics[height=0.91\linewidth]{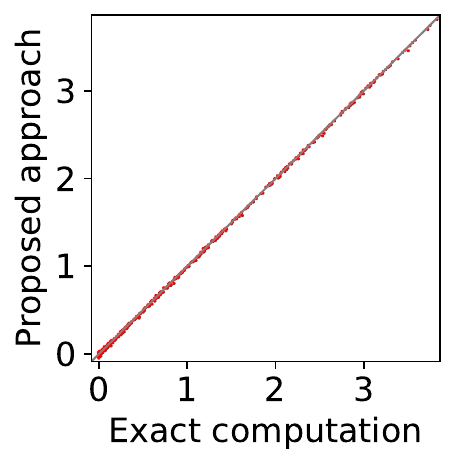}
    (a) players' utility
  \end{minipage}
  \begin{minipage}{0.24\linewidth}
    \centering
    \includegraphics[height=0.91\linewidth]{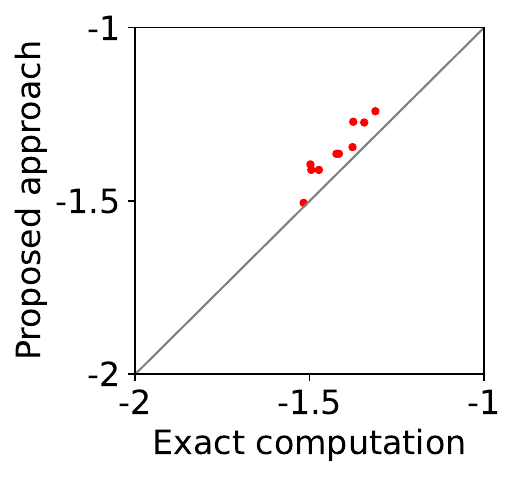}
    (b) mediator's revenue
  \end{minipage}
  \begin{minipage}{0.02\linewidth}
    \
  \end{minipage}
  \begin{minipage}{0.24\linewidth}
    \centering
    \includegraphics[height=0.91\linewidth]{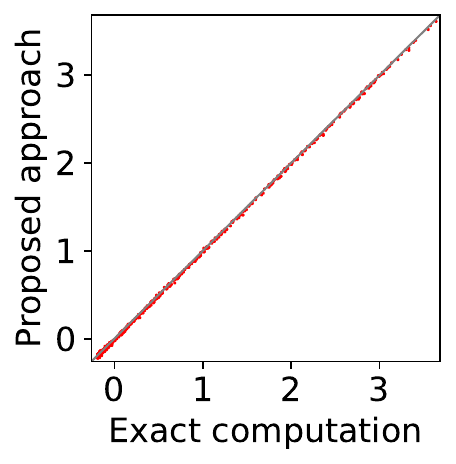}
    (c) players' utility
  \end{minipage}
  \begin{minipage}{0.24\linewidth}
    \centering
    \includegraphics[height=0.91\linewidth]{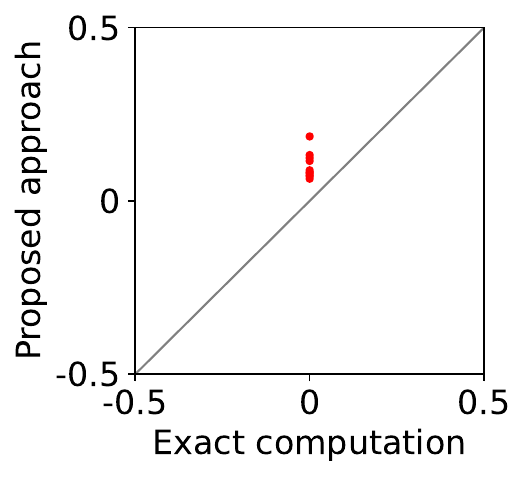}
    (d) mediator's revenue
  \end{minipage}
  \caption{The red dots show the expected utility of the players in Columns (a) and (c) and the expected revenue of the mediator in Columns (b) and (d), where analytical solutions are evaluated exactly (horizontal axes) or estimated with $(0.25,0.1)$-PAC SE-BME (vertical axes) for environments with $|\cN|=8$ players, each having $|\cT_n|=8$ possible types.  The analytical solution guarantees $\theta$-IR with $\theta\equiv 0$ in Columns (a) and (b) and \underline{$\rho'$-SBB with $\rho'=0.1$ in Columns (c) and (d)}.  Results are plotted for 10 random seeds.  Diagonal lines are also plotted to help understand where the horizontal and vertical axes are equal.}
  \label{fig:bb-ir-rho}
\end{figure}

As an example, we set $\rho'=0.1$ in Figure~\ref{fig:bb-ir-rho}.  The consequence of replacing $\rho=0$ with $\rho'=0.1$ is as expected.  In Columns (b) and (d), the expected revenue of the mediator when the best mean is estimated with BME (Proposed approach) is shifted to the above by $\rho'-\rho=0.1$.  Although it may be unclear from Columns (a) and (c), the corresponding expected utility of each player is shifted to the left by $(\rho'-\rho)/|\cN|=0.0125$.  In practice, we may choose $\rho'$ and $\theta'$ by taking into account these shifts as well as the condition on the feasibility of LP (Lemma~\ref{lemma:condition}).

\begin{figure}
  \begin{minipage}{0.49\linewidth}
    \centering
    \includegraphics[width=\linewidth]{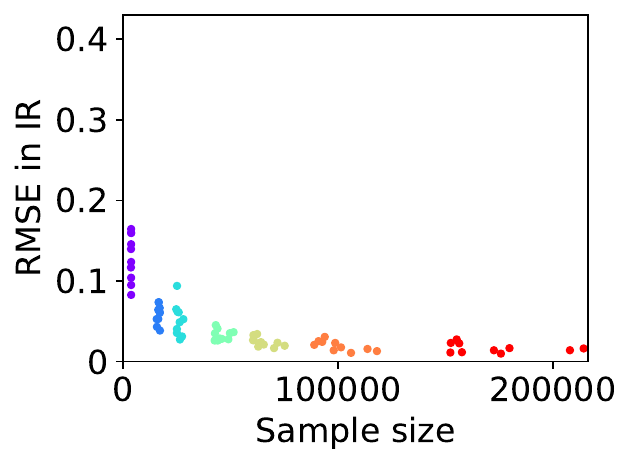}
    (a) players' utility
  \end{minipage}
  \begin{minipage}{0.49\linewidth}
    \centering
    \includegraphics[width=\linewidth]{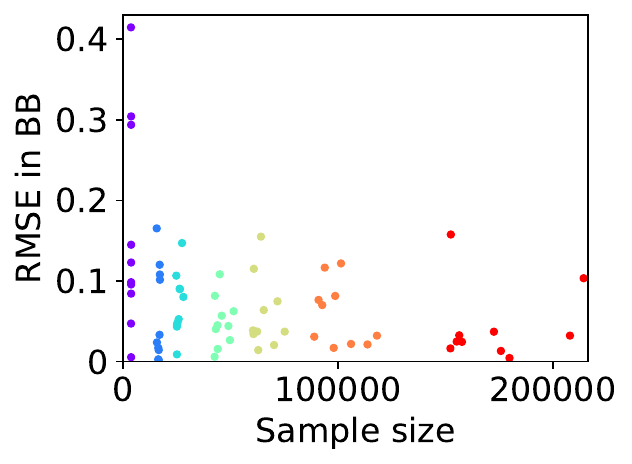}
    (b) mediator's revenue
  \end{minipage}
  \caption{Root mean squared error in (a) the expected utility of each player and (b) the expected revenue of the mediator against the total sample size, when there are $|\cN|=8$ players, each with $|\cT_n|=4$ possible types.  Here, we set $\delta=0.1$ and vary $\varepsilon$ from 1.0 (purple), 0.5, 0.4, 0.3, 0.25, 0.2, to 0.15 (red).}
  \label{fig:rmse}
\end{figure}

In Figure~\ref{fig:rmse}, we show the Root Mean Squared Error (RMSE) in the expected utility of each player (a) and the expected revenue of the mediator (b) that are estimated with BME for the case with $|\cN|=8$ players, each with $|\cT_n|=4$ types.  Here, we fix $\delta=0.1$ and vary $\varepsilon$ from 1.0 to 0.15 in the BME.  For each pair of $(\delta,\varepsilon)$, the experiments are repeated 10 times with different random seeds.  The total sample size increases as the value of $\varepsilon$ decreases.  Hence, the purple dots correspond to $\varepsilon=1.0$, and the red dots are $\varepsilon=0.15$.  Overall, we can observe that RMSE can be reduced by using small $\varepsilon$ at the expense of increased sample size, that relatively large values such as $\varepsilon=0.5$ gives reasonably small RMSE, and that larger values of $\varepsilon$ have diminishing effects on RMSE.

\subsubsection{Computational requirements}
\label{sec:detail:exp:require}

We have run all the experiments on a single core with at most 66~GB memory without GPUs in a cloud environment.  The associated source code is submitted as a supplementary material and will be open-sourced upon acceptance.  Table~\ref{tbl:comp} summarizes the CPU time and maximum memory require to generate each figure.  For example, CPU time for Figure~\ref{fig:BME-BAI}(a) is the time to generate three panels in Column (a) of Figure~\ref{fig:BME-BAI}.  Note that the CPU time and maximum memory reported in Table~\ref{tbl:comp} are not optimized and include time and memory for storing intermediate results and other processing for debugging purposes; these should be understood as the computational requirements to execute the source code as is.

\begin{threeparttable}
  \caption{CPU time and maximum memory required to generate figures}
  \label{tbl:comp}
  \centering
  \begin{tabular}{lrr}
    \toprule
    Figure   & CPU Time (seconds) & Max Memory (GB) \\
    \midrule
    Figure~\ref{fig:BME-BAI}(a) & 413.1 & $<1$ \\
    Figure~\ref{fig:BME-BAI}(b) & 81.9 & $<1$ \\
    Figure~\ref{fig:BME-BAI}(c) & 40.4 & $<1$ \\
    Figure~\ref{fig:sample}(a)\tnote{\textdagger} & 0.7 & $<1$ \\
    Figure~\ref{fig:sample}(b)\tnote{\textdagger} & 1.1 & $<1$ \\
    Figure~\ref{fig:sample}(c)\tnote{\textdagger} & 80.1 & 1.9 \\
    Figure~\ref{fig:n-eval}(a) and Figure~\ref{fig:n-eval-total}(a) & 7,883.0 & 65.5 \\
    Figure~\ref{fig:n-eval}(b) and Figure~\ref{fig:n-eval-total}(b) & 2,935.2 & 17.0 \\
    Figure~\ref{fig:bb-ir}(a)-(b) and Figure~\ref{fig:bb-ir-rho}(a)-(b)\tnote{\textdaggerdbl} & 17,531.6 & 1.6 \\
    Figure~\ref{fig:bb-ir}(c)-(d) and Figure~\ref{fig:bb-ir-rho}(c)-(d)\tnote{\textdaggerdbl} & 17,416.3 & 1.6 \\
    Figure~\ref{fig:rmse} & 527.9 & $<1$ \\
    \bottomrule
  \end{tabular}
  \begin{tablenotes}
  \item[\textdagger] Figure~\ref{fig:sample} shows the results with one random seed, but here the CPU Time reports the average over 10 seeds, and Max Memory reports the maximum over 10 seeds.
  \item[\textdaggerdbl] Figure~\ref{fig:bb-ir-rho} could have been obtained by simply reusing and shifting Figure~\ref{fig:bb-ir}, but here the CPU time reports the time to generate the two figures without reuse.
  \end{tablenotes}
\end{threeparttable}

\subsection{Details of Section~\ref{sec:conclusion}}
\label{sec:detail:conclusion}

\subsubsection{Limitations}
\label{sec:detail:conclusion:limit}

While the proposed approach makes major advancement in the field, it certainly has limitations.  Here, we discuss four major limitations of this work as well as interesting directions of research motivated by those limitations.

First, when types are not independent between players, the sufficient condition in \eqref{eq:condition} may not be necessary (Lemma~\ref{lemma:condition}).  This means that the LP may be feasible even when the condition in the lemma is violated, and our results do not provide optimal solutions for those cases.  Further research is needed to understand exactly when the sufficient condition is also necessary.  It is also important to develop efficient methods for solving the LP when the types are dependent.

Second, our mechanisms guarantee strong budget balance (SBB) and individual rationality (IR) in expectation with respect to the distribution of the players' types, but this does not guarantee that those properties are satisfied \textit{ex post} (for any realization of the types).  Although the satisfaction in expectation is often sufficient for risk-neutral decision makers, it is important to let the mediator and the participants aware that they may experience negative utilities even if their expected utilities are nonnegative.  It would also be an interesting direction of research to extend the proposed approach towards achieving these properties \textit{ex post}.

Finally, our experiments have considered environments with up to 128 players, each with at most 16 types.  Although these are substantially larger than the environment studied in \cite{osogami2023learning}, they certainly do not cover the scale needed for all applications.  While (10-100 times) larger environments could be handled with improved implementation and greater computational resources, essentially new ideas would be needed for substantially (over $10^3$ times) larger environments or continuous type space.  It would be an interesting direction of research to identify and exploit structures of particular environments for designing scalable approaches of mechanism design for those environments.

\subsubsection{Societal impacts}
\label{sec:detail:conclusion:impact}

We expect that the proposed approach has several positive impacts on trading networks in particular and the society in general.  In particular, the proposed approach enables mechanisms that can maximize the efficiency of a trading network and minimize the fees that the participants need to pay to the mediator.  Also, the DSIC guaranteed by the proposed approach would make it more difficult for malicious participants to manipulate the outcome of a trading network.

On the other hand, the proposed approach might have negative impacts depending on where and how it is applied.  For example, although the proposed approach guarantees individual rationality, some participants might benefit less from the mechanism designed with our approach than other participants.  This can happen, because maximizing the social welfare does not mean that all the participants are treated fairly.  Before applying the mechanisms designed with the proposed approach, it is thus recommended assessing whether such fairness needs to be considered and to take any actions that mitigate the bias if needed.

\section{Proofs}
\label{sec:proof}

\subsection{Proofs of the lemmas and corollaries in Section~\ref{sec:solution}}

\begin{proof}[Proof of Lemma~\ref{lemma:condition}]
  We start by establishing the sufficiency of \eqref{eq:condition} for the feasibility of the LP \eqref{eq:LP-obj}-\eqref{eq:LP-WBB}, regardless of whether the types are independent or not.

  We will show that
  \begin{align}
    h_n(t_{-n})
    & = \eta_n
    \coloneqq \min_{t_n\in\cT_n} \left\{
    \bE[w^\star(t) \mid t_n] - \theta(t_n)
    \right\}
    \qquad\forall t_{-n}\in\cT_{-n}
    \label{eq:solution}
  \end{align}
  is a feasible solution when \eqref{eq:condition} holds.

  The $\theta$-IR \eqref{eq:LP-IR} is satisfied with \eqref{eq:solution}, because for any $n\in\cN$ we have
  \begin{align}
    \bE[h_n(t_{-n}) \mid t_n]
    & = \eta_n \\
    & = \min_{t_n'\in\cT_n} \left\{
    \bE[w^\star(t') \mid t_n'] - \theta(t_n')
    \right\} \\
    & \le \bE[w^\star(t) \mid t_n] - \theta(t_n).
  \end{align}

  The $\rho$-WBB \eqref{eq:LP-WBB} is satisfied with \eqref{eq:solution}, because
  \begin{align}
    \sum_{n\in\cN} \bE[h_n(t_{-n})]
    & = \sum_{n\in\cN} \eta_n \\
    & = \sum_{n\in\cN} \min_{t_n\in\cT_n} \left\{
    \bE[w^\star(t) \mid t_n] - \theta(t_n)
    \right\} \\
    & \ge (N - 1) \, \bE[w^\star(t)] + \rho,
  \end{align}
  where the inequality follows from \eqref{eq:condition}.
  This establishes the sufficiency of \eqref{eq:condition}.

  Next, we prove the necessity of \eqref{eq:condition} for the feasibility of the LP \eqref{eq:LP-obj}-\eqref{eq:LP-WBB} when the types are independent.
  When the types are independent, \eqref{eq:LP-IR} is reduced to
  \begin{align}
    & \bE[w^\star(t) \mid t_n] - \bE[h_n(t_{-n})]
    \ge \theta(t_n)
    \qquad \forall t_n\in\cT_n, \forall n\in\cN \\
    & \iff
    \min_{t_n\in\cT_n} \left\{
    \bE[w^\star(t) \mid t_n] - \theta(t_n)
    \right\}
    \ge \bE[h_n(t_{-n})]
    \qquad \forall n\in\cN.
  \end{align}
  This together with \eqref{eq:LP-WBB} establishes the necessity of
  \begin{align}
    (N - 1) \, \bE[w^\star(t)] + \rho
    & \le \sum_{n\in\cN} \bE[h_n(t_{-n})] \\
    & \le \sum_{n\in\cN} \min_{t_n\in\cT_n} \left\{
    \bE[w^\star(t) \mid t_n] - \theta(t_n)
    \right\}.
  \end{align}

  Finally, we construct an example that satisfies \eqref{eq:LP-IR}-\eqref{eq:LP-WBB} but violates \eqref{eq:condition}.  Let $\cN=\{1,2\}$; $\cT_n=\cT\coloneqq\{1,2\}, \forall n\in\cN$; $\rho=0$; $\theta(m)=0, \forall m\in\cT$.  We assume that the types are completely dependent (namely, $t_1=t_2$ surely) and let $p$ be the probability that $t_1=t_2=1$ (hence, $t_1=t_2=2$ with probability $1-p$).

    For this example, we rewrite \eqref{eq:LP-IR}-\eqref{eq:LP-WBB} and \eqref{eq:condition} by using $x_m\coloneqq w^\star((m,m))$ and $y_{nm}\coloneqq h_n(m)$ for $m\in\cT$ and $n\in\cN$.  Notice that, for any $m\in\cT$ and $n\in\cN$, we have
    \begin{align}
        \bE[w^\star(t)\mid t_n=m] & = x_m \\
        \bE[h_n(t_{-n})\mid t_n=m] & = y_{nm},
    \end{align}
    since types are completely dependent.  Hence, \eqref{eq:LP-IR} is reduced to
    \begin{align}
        x_m - y_{nm} \ge 0 \qquad \forall m\in\cT, \forall n\in\cN
        \label{eq:unneeded-IR}
    \end{align}
    and \eqref{eq:LP-WBB} is reduced to
    \begin{align}
        p \, (y_{11}+y_{21}-x_1)
        + (1-p) \, (y_{12}+y_{22}-x_2) \ge 0.
        \label{eq:unneeded-WBB}
    \end{align}
    On the other hand, \eqref{eq:condition} is reduced to
    \begin{align}
        2 \,\min\{x_1, x_2\} \ge p \, x_1 + (1-p) \, x_2.
        \label{eq:unneeded-condition}
    \end{align}

    Consider the case where $x_m > 0, \forall m\in\cT$.  In this case, \eqref{eq:unneeded-IR}-\eqref{eq:unneeded-WBB} suggest that \eqref{eq:LP-IR}-\eqref{eq:LP-WBB} are satisfied as long as $y_{nm}$ satisfies
    \begin{align}
        \frac{x_m}{2} \le y_{nm} \le x_m \qquad \forall m\in\cT, \forall n\in\cN,
        \label{eq:unneeded-IR-WBB}
    \end{align}
    whether \eqref{eq:unneeded-condition} is satisfied or not.  Indeed, \eqref{eq:unneeded-IR-WBB} can be met even if \eqref{eq:unneeded-condition} is violated, for example when $p=\frac{1}{2}, x_1=1, x_2=4, y_{nm}=\frac{2}{3}x_m, \forall m\in\cT, \forall n\in\cN$; this serves as a desired example, concluding the proof.
\end{proof}

\begin{proof}[Proof of Lemma~\ref{lemma:optimal}]
We first rewrite the LP \eqref{eq:LP-obj}-\eqref{eq:LP-WBB} in the following equivalent form:
\begin{align}
  \min_h
  & \qquad\sum_{n\in\cN} \sum_{t_n\in\cT_n} \bP[t_n] \, \bE[h_n(t_{-n}) \mid t_n] & \label{eq:LP-obj-equiv}\\
  \mathrm{s.t.}
  & \qquad \bE[w^\star(t) \mid t_n] - \bE[h_n(t_{-n}) \mid t_n]
  \ge \theta(t_n)
  \qquad \forall t_n\in\cT_n, \forall n\in\cN
  \label{eq:LP-IR-equiv}\\
  & \qquad\sum_{n\in\cN} \sum_{t_n\in\cT_n} \bP[t_n] \, \bE[h_n(t_{-n}) \mid t_n]
  - (N - 1) \, \bE[w^\star(t)]
  \ge \rho.
  \label{eq:LP-WBB-equiv}
\end{align}
Then it can be easily observed that the optimal objective value must be equal to $(N - 1) \, \bE[w^\star(t)] + \rho$ (i.e., when equality holds in \eqref{eq:LP-WBB-equiv}), since changing $h$ in a way it decreases the value of $\bE[h_n(t_{-n}) \mid t_n]$ only makes \eqref{eq:LP-IR-equiv} more satisfiable\footnote{This implies that $\rho$-SBB is satisfied whenever $\rho$-WBB is satisfied}.

Hence, to prove that any $h\in\cH$ is an optimal solution, it suffices to show that \eqref{eq:LP-WBB-equiv} is satisfied with equality and \eqref{eq:LP-IR-equiv} is satisfied with any $h\in\cH$.  When $h\in\cH$, we have, for any $t_n$, that
\begin{align}
    \bE[h_n(t_{-n}) \mid t_n]
    & = \eta_n \\
    & \le \min_{t_n'\in\cT_n} \left\{
    \bE[w^\star(t) \mid t_n']  - \theta(t_n')
    \right\}\\
    & \le \bE[w^\star(t) \mid t_n]  - \theta(t_n),
\end{align}
where the first inequality follows from \eqref{eq:simplex1}.  We also have
\begin{align}
    \sum_{n\in\cN} \sum_{t_n\in\cT_n} \bP[t_n] \, \bE[h_n(t_{-n}) \mid t_n]
    & = \sum_{n\in\cN} \sum_{t_n\in\cT_n} \bP[t_n] \, \eta_n \\
    & = \sum_{n\in\cN} \eta_n \\
    & = (N - 1) \, \bE[w^\star(t)] + \rho,
\end{align}
where the last equality follows from \eqref{eq:eta-n} and \eqref{eq:simplex2}.

Finally, when \eqref{eq:condition} holds, $\cH$ is nonempty, because the following $\eta_n$ satisfies the conditions \eqref{eq:eta-n}-\eqref{eq:simplex2}:
\begin{align}
    \eta_n
    & = \min_{t_n\in\cT_n} \left\{
    \bE[w^\star(t) \mid t_n]  - \theta(t_n)
    \right\} - \delta
    \label{eq:optimal-h}
\intertext{where}
    \delta
    & \coloneqq \frac{1}{N} \bigg(
    \sum_{n\in\cN} \min_{t_n\in\cT_n} \left\{
    \bE[w^\star(t) \mid t_n]  - \theta(t_n)
    \right\}
    - (N - 1) \, \bE[w^\star(t)] - \rho
    \bigg).
    \label{eq:optimal-delta}
\end{align}
Notice that $\delta\ge 0$ follows from \eqref{eq:condition}.

This establishes the sufficiency of \eqref{eq:condition} for an optimal solution to exist in $\cH$.
The necessity follows in exactly the same way as the proof of the necessary condition in Lemma~\ref{lemma:condition}.
\end{proof}

\begin{corollary}
    Any VCG mechanism given with a pivot rule in $\cH$ satisfies $\rho$-SBB.
    \label{cor:BB}
\end{corollary}
\begin{proof}[Proof of Corollary~\ref{cor:BB}]
    By \eqref{eq:eta-n}, we have
    \begin{align}
        \lefteqn{\sum_{n\in\cN} \eta_n - (N-1) \, \bE[w^\star(t)] - \rho} \notag\\
        & = \sum_{n\in\cN} \min_{t_n\in\cT_n} \left\{\bE[w^\star(t)\mid t_n] - \theta(t_n)\right\}
        - \sum_{n\in\cN} \delta_n - (N-1) \, \bE[w^\star(t)] - \rho \\
        & = 0,
    \end{align}
    where the last equality follows from \eqref{eq:simplex2}.  Hence, \eqref{eq:LP-WBB} is satisfied with equality.
\end{proof}

\begin{corollary}
    Any VCG mechanism with a pivot rule that satisfies \eqref{eq:eta-n} and \eqref{eq:simplex2} ensures $\rho$-SBB.
    Any VCG mechanism with a pivot rule that satisfies \eqref{eq:eta-n} and \eqref{eq:simplex1} ensures $\theta$-IR for any $t_n\in\cT_n$ and $n\in\cN$.
    \label{cor:SBB-IR}
\end{corollary}
\begin{proof}[Proof of Corollary~\ref{cor:SBB-IR}]
  The first part of the corollary can be proved analogously to Corollary~\ref{cor:BB}.
  Regarding the second part of the corollary, by \eqref{eq:eta-n}, for any $t_n\in\cT_n$ and $n\in\cN$, we have
    \begin{align}
        \lefteqn{\bE[w^\star(t)\mid t_n] - \eta_n - \theta(t_n)} \notag\\
        & = \bE[w^\star(t)\mid t_n] - \theta(t_n) - \left(
        \min_{t_n'\in\cT_n} \left\{\bE[w^\star(t)\mid t_n'] - \theta(t_n')\right\}
        \right) + \delta_n,
    \end{align}
    which is nonnegative by \eqref{eq:simplex1}, and hence \eqref{eq:LP-IR} holds.
\end{proof}

\subsection{Proofs of the lemmas in Section~\ref{sec:bandit2} and Appendix~\ref{sec:detail:bandit2}}
\label{sec:proof:bandit2}

\begin{proof}[Proof of Lemma~\ref{lemma:bai2bme}]
Since the sample complexity of \texttt{PAC-BAI} in Step~\ref{step:best-reward:bai} is $M$ and Step~\ref{step:best-reward:pull} pulls an arm $m^\star$ times, the sample complexity of Algorithm~\ref{alg:best-reward} is $M+m^\star$.  Hence, it remains to prove \eqref{eq:toshow-upperbound}.

Recall that $\hat I$ is a random variable representing the index of the best arm returned by an $(\varepsilon,\delta)$-PAC BAI.  Then we have the following bound:
    \begin{align}
    \lefteqn{
        \Pr\left( |\hat\mu_{\hat I} - \mu_\star| > \frac{3}{2}\varepsilon \right)
    } \notag\\
    & = \Pr\left(\hat\mu_{\hat I} > \mu_\star + \frac{3}{2}\varepsilon\right)
    +
    \Pr\left(\hat\mu_{\hat I} < \mu_\star - \frac{3}{2}\varepsilon\right)\\
    & \le \Pr\left(\hat\mu_{\hat I} > \mu_{\hat I} + \frac{3}{2}\varepsilon\right) \notag\\
    & \hspace{3mm} + \Pr\left(
    \left\{
        \hat\mu_{\hat I} < \mu_\star - \frac{3}{2}\varepsilon
    \right\}
    \cap
    \left\{
        \mu_{\hat I} < \mu_\star - \varepsilon
    \right\}
    \right) + \Pr\left(
    \left\{
        \hat\mu_{\hat I} < \mu_\star - \frac{3}{2}\varepsilon
    \right\}
    \cap
    \left\{
        \mu_{\hat I} \ge \mu_\star - \varepsilon
    \right\}
    \right) \\
    & \le \Pr\left(\hat\mu_{\hat I} > \mu_{\hat I} + \frac{3}{2}\varepsilon\right)
    + \Pr\left( \mu_{\hat I} < \mu_\star - \varepsilon \right)
    + \Pr\left(
    \left\{
        \hat\mu_{\hat I} < \mu_\star - \frac{3}{2}\varepsilon
    \right\}
    \cap
    \left\{
        \mu_{\hat I} \ge \mu_\star - \varepsilon
    \right\}
    \right) \\
    & \le \Pr\left(\hat\mu_{\hat I} > \mu_{\hat I} + \frac{3}{2}\varepsilon\right)
    + \Pr\left( \mu_{\hat I} < \mu_\star - \varepsilon \right)
    + \Pr\left( \hat\mu_{\hat I} < \mu_{\hat I} - \frac{1}{2}\varepsilon \right) \\
    & \le \Pr\left(\hat\mu_{\hat I} > \mu_{\hat I} + \frac{3}{2}\varepsilon\right)
    + \delta
    + \Pr\left(\hat\mu_{\hat I} < \mu_{\hat I} - \frac{1}{2}\varepsilon\right),
    \label{eq:toapplyhoeffding}
\end{align}
where the last inequality follows from PAC($\varepsilon,\delta$) of BAI.

Since $\hat\mu_{\hat I}$ is the average of $m^\star$ samples from arm $\hat I$, applying Hoeffding's inequality to the last term of \eqref{eq:toapplyhoeffding}, we obtain
\begin{align}
    \Pr\left( \hat\mu_{\hat I} < \mu_{\hat I} - \frac{1}{2} \varepsilon \right)
    & = \sum_{k\in[1,K]} \Pr\left( \hat\mu_k < \mu_k - \frac{1}{2} \varepsilon \midd \hat I=k \right) \, \Pr(\hat I=k) \\
    & \le \sum_{k\in[1,K]} \exp\left( -2 \left(\frac{1}{2}\varepsilon \right)^2 m^\star \right) \, \Pr(\hat I=k) \\
    & = \exp\left( -2 \left(\frac{1}{2}\varepsilon \right)^2 m^\star \right)\label{eq:detail1},
    \intertext{where the inequality is obtained by applying Hoeffding's inequality to the sample mean $\hat\mu_k$ of $m^\star$ independent random variables having support in $[0,1]$.  We can also show the following inequality in an analogous manner:}
    \Pr\left( \hat\mu_{\hat I} > \mu_{\hat I} + \frac{3}{2} \varepsilon \right)
    & \le \exp\left( -2 \left(\frac{3}{2}\varepsilon \right)^2 m^\star \right).\label{eq:detail2}
\end{align}
By applying \eqref{eq:detail1}-\eqref{eq:detail2} to \eqref{eq:toapplyhoeffding}, we finally establish the bound to be shown:
\begin{align}
        \Pr\left( |\hat\mu_{\hat I} - \mu_\star| > \frac{3}{2}\varepsilon \right)
    & \le \delta
    + \exp\left(-2\,\left(\frac{3}{2}\varepsilon\right)^2\,m^\star\right)
    + \exp\left(-2\,\left(\frac{1}{2}\varepsilon\right)^2\,m^\star\right) \\
    & \le \delta
    + \exp\left(-2\,\left(\frac{3}{2}\varepsilon\right)^2\,\frac{2}{\varepsilon^2} \log \frac{1.22}{\delta}\right)
    + \exp\left(-2\,\left(\frac{1}{2}\varepsilon\right)^2\,\frac{2}{\varepsilon^2} \log \frac{1.22}{\delta}\right) \notag\\
    & \qquad\mbox{by the definition of $m^\star$}\\
    & = \delta
    + \left(\frac{\delta}{1.22}\right)^9
    + \frac{\delta}{1.22}\\
    & \le \left(1 + \frac{1}{1.22^9} + \frac{1}{1.22}\right) \delta \\
    & \le 2 \, \delta.
\end{align}
\end{proof}

\begin{proof}[Proof of Lemma~\ref{lemma:bias}]
    Although the lemma is stated in \cite{evendar02pac} with reference to \cite{chernoff1972sequential}, this specific lemma is neither stated nor proved explicitly in \cite{chernoff1972sequential}.  For completeness, here, we prove the lemma following the general methodology provided in \cite{chernoff1972sequential}.  Specifically, we derive the expected sample size required by the sequential probability-ratio test (SPRT; Section 10 of \cite{chernoff1972sequential}), whose optimality (Theorem 12.1 of \cite{chernoff1972sequential}) will then establish the lemma.

    Consider two hypotheses, $\theta_1$ and $\theta_2$, for the probability distribution $\bP$ of a random variable $X$, which takes either the value of 1 or $-1$, where
    \begin{align}
        \bP(X=1\mid \theta_1) & = \frac{1+\varepsilon}{2}\\
        \bP(X=1\mid \theta_2) & = \frac{1-\varepsilon}{2}
    \end{align}
    Consider the SPRT procedure that takes i.i.d.\ samples, $X_1, X_2, \ldots, X_N$, from $\bP$ until the stopping time $N$ when
    \begin{align}
        \lambda_N
        & \coloneqq \prod_{n=1}^N \frac{\bP(X_n\mid \theta_1)}{\bP(X_n\mid\theta_2)}
        = \prod_{n=1}^N \left(\frac{\varepsilon+1}{\varepsilon-1}\right)^{X_n}
    \end{align}
    hits either $A\in\bR$ or $1/A$.  When $\lambda_N$ hits $A$, we identify $\theta_1$ as the correct hypothesis.  When $\lambda_N$ hits $1/A$, we identify $\theta_2$ as the correct hypothesis.

    Let
    \begin{align}
        S_N
        \coloneqq \log \lambda_N
        = \sum_{n=1}^N X_n \, \log\frac{1+\varepsilon}{1-\varepsilon}.
    \end{align}
    Since $N$ is a stopping time, by Wald's lemma, we have
    \begin{align}
        \bE[S_N\mid\theta_1]
        & = \bE[N \mid \theta_1] \, \bE[X\mid \theta_1] \, \log\frac{1+\varepsilon}{1-\varepsilon} \label{eq:wald1}\\
        & = \bE[N \mid \theta_1] \, \varepsilon \, \log\frac{1+\varepsilon}{1-\varepsilon} \\
        \bE[S_N\mid\theta_2]
        & = - \bE[N \mid \theta_2] \, \varepsilon \, \log\frac{1+\varepsilon}{1-\varepsilon}.
    \end{align}
    Let $\delta$ be the probability of making the error in identifying the correct hypothesis.  Then we must have
    \begin{align}
        \bE[S_N\mid\theta_1]
        & = (1-\delta) \, \log A + \delta \, \log(1/A) \\
        \bE[S_N\mid\theta_2]
        & = \delta \, \log A + (1 - \delta) \, \log(1/A). \label{eq:wald4}
    \end{align}
    By \eqref{eq:wald1}-\eqref{eq:wald4}, we have
    \begin{align}
        \bE[N\mid\theta_1]
        = \bE[N\mid\theta_2]
        = \frac{1-2\,\delta}{\varepsilon\log\frac{1+\varepsilon}{1-\varepsilon}} \, \log A. \label{eq:EN}
    \end{align}

    Now, notice that $S_N$ hits $\log A$ when we have
    \begin{align}
        |\{n: X_n=1\}| - |\{n: X_n=-1\}| \ge
        \frac{\log A}{\log\frac{1+\varepsilon}{1-\varepsilon}}
    \end{align}
    for the first time and hits $-\log A$ when we have
    \begin{align}
        |\{n: X_n=-1\}| - |\{n: X_n=1\}| \ge
        \frac{\log A}{\log\frac{1+\varepsilon}{1-\varepsilon}}
    \end{align}
    for the first time.  Hence, by the gambler's ruin probability, we have
    \begin{align}
        \delta
        & = \frac{1-\left(\frac{1+\varepsilon}{1-\varepsilon}\right)^{\frac{\log A}{\log\frac{1+\varepsilon}{1-\varepsilon}}}}{1-\left(\frac{1+\varepsilon}{1-\varepsilon}\right)^{2\frac{\log A}{\log\frac{1+\varepsilon}{1-\varepsilon}}}}
        = \frac{1}{1+\left(\frac{1+\varepsilon}{1-\varepsilon}\right)^{\frac{\log A}{\log\frac{1+\varepsilon}{1-\varepsilon}}}},
    \end{align}
    which implies
    \begin{align}
        A & = \frac{1-\delta}{\delta}.
    \end{align}

    Plugging the last expression into \eqref{eq:EN}, we obtain
    \begin{align}
        \bE[N\mid\theta_1]
        = \bE[N\mid\theta_2]
        = \frac{1-2\,\delta}{\varepsilon\log\frac{1+\varepsilon}{1-\varepsilon}} \, \log \frac{1-\delta}{\delta}
        = O\left(\frac{1}{\varepsilon^2}\log \frac{1}{\delta}\right).
    \end{align}
\end{proof}

\begin{proof}[Proof of Lemma~\ref{lemma:lower}]
    We will construct an algorithm that correctly identifies the mean $\alpha$ of $B$ with probability at least $\delta$ using at most $M/K$ samples of $B$ in expectation given the access to an $(\varepsilon/2,\delta/2)$-PAC BME with sample complexity $M$.  First, the algorithm independently draws $i^+$ and $i^-$ from the uniform distribution over $[1, K]$.

    Consider two environments of $K$-armed bandit, $\cE^+$ and $\cE^-$, where every arm gives the reward according to the Bernoulli distribution with mean $\alpha^-=(1-\varepsilon)/2$ except that the reward of arm $i^+$ in $\cE^+$ has the same distribution as $B$ and the reward of arm $i^-$ in $\cE^-$ has the same distribution as $1-B$.  Note that the algorithm can simulate the reward with the known mean $\alpha^-$ using the algorithm's internal random number generators.  Only when the algorithm pulls arm $i^+$ of $\cE^+$ or arm $i^-$ of $\cE^-$, it uses the sample of $B$, which contributes to the sample complexity of the algorithm.

    The algorithm then runs two copies of the $(\varepsilon/2,\delta/2)$-PAC BME with sample complexity $M$ in parallel: one referred to as BME$^+$ is run on $\cT^+$, and the other referred to as BME$^-$ is run on $\cT^-$.  At each step, let $M^+$ be the number of samples BME$^+$ has taken from arm $i^+$ by that step and $M^-$ be the corresponding number BME$^-$ has taken from arm $i^-$.  If $M^+<M^-$, the algorithm lets  BME$^+$ pull an arm; otherwise, the algorithm lets BME$^-$ pull an arm.  Therefore, $|M^+-M^-|\le 1$ at any step.

    This process is continued until one of BME$^+$ and BME$^-$ terminates and returns an estimate $\hat\mu$ of the best mean.  If BME$^+$ terminates first, then the algorithm determines that $\alpha=\alpha^-$ if $\hat\mu<1/2$ and that $\alpha=\alpha^+$ otherwise.  If BME$^-$ terminates first, then the algorithm determines that $\alpha=\alpha^+$ if $\hat\mu<1/2$ and that $\alpha=\alpha^-$ otherwise.  Due to the $(\varepsilon/2,\delta/2)$-PAC property of BME$^+$ and BME$^-$, the algorithm correctly identifies the mean of $B$ with probability at least $1-\delta$.  Formally, if $\alpha=\alpha^+$, we have
    \begin{align}
        \Pr\left(\hat\mu<\frac{1}{2}\right)
        & =
        \Pr\left(\hat\mu<\frac{1}{2} \midd \mbox{BME}^+ \mbox{ terminates first} \right) \, \Pr(\mbox{BME}^+ \mbox{ terminates first}) \notag\\
        & \hspace{3mm} +
        \Pr\left(\hat\mu<\frac{1}{2} \midd \mbox{BME}^- \mbox{ terminates first} \right) \, \Pr(\mbox{BME}^- \mbox{ terminates first}) \\
        & \le \frac{\delta}{2} + \frac{\delta}{2} \\
        & = \delta.
    \end{align}
    Analogously, $\Pr\left(\hat\mu<\frac{1}{2}\right)\le\delta$ can be shown if $\alpha=\alpha^-$.

    What remains to prove is the sample complexity of the algorithm.  Recall that each of BME$^+$ and BME$^-$ pulls arms at most $M$ times before it terminates due to their sample complexity.  Notice that the arms in $\cE^-$ are indistinguishable when $\alpha=\alpha^-$, and the arms in $\cE^+$ are indistinguishable when $\alpha=\alpha^+$.  Therefore, at least one of BME$^+$ and BME$^-$ is run on the environment where the arms are indistinguishable.  Since $i^-$ and $i^-$ are sampled uniformly at random from $[1,K]$, BME (either BME$^+$ or BME$^-$) would take at most $M/n$ samples from $B$ in expectation if the arms are indistinguishable.  Since $|M^+-M^-|\le 1$, we establish that the sample complexity of the algorithm is $O(M/n)$ in expectation.
\end{proof}

\subsection{Proofs of the lemma, theorem, and proposition in Section~\ref{sec:bandit}}

\begin{proof}[Proof of Lemma~\ref{lemma:connect}]
    Note that the PAC estimators can give independent estimates, $\tilde\kappa_n(\theta)$ for $n\in\cN$ and $\tilde\lambda$, such that
    \begin{align}
        \Pr\left(|\tilde\kappa_n(\theta)-\kappa_n(\theta)|\le\varepsilon'\right)
        & \ge 1 - \delta' \qquad\forall n\in\cN \\
        \Pr\left(|\tilde\lambda(\theta)-\lambda(\theta)|\le\varepsilon''\right) & \ge 1 - \delta',
    \end{align}
    which imply
    \begin{align}
        \Pr\left(
            |\tilde\lambda(\theta)-\lambda(\theta)|\le\varepsilon'',
            |\tilde\kappa_n(\theta)-\kappa_n(\theta)|\le\varepsilon', \forall n\in\cN
        \right)
        & \ge (1 - \delta')^{N+1}.
    \end{align}
    Hence, with probability at least $(1 - \delta')^{N+1}$, the expected utility of player $n$ given its type $t_n$ (recall \eqref{eq:IRequiv}) is
    \begin{align}
        \E[w^\star(t)\mid t_n] - (\tilde\kappa_n(\theta)-\tilde d_n)
        & \ge \E[w^\star(t)\mid t_n] - \kappa_n(\theta) - \varepsilon' + \varepsilon''' \\
        & \ge \theta(t_n) - (\varepsilon' - \varepsilon'''),
    \end{align}
    where the first inequality follows from the PAC bound on $\tilde\kappa_n(\theta)$ and the definition of $\tilde d_n$, and the last inequality follows from the original guarantee when $\kappa_n(\theta)$ is exactly computed, and the expected revenue of the mediator (recall \eqref{eq:WBBequiv}) is
    \begin{align}
        \sum_{n\in\cN} (\tilde\kappa_n(\theta) - \tilde d_n) - (N-1) \, \E[w^\star(t)]
        & = (N-1) \, (\tilde\lambda + \varepsilon''') - (N-1) \, \E[w^\star(t)] \\
        & \ge \rho + (N-1) \, (\varepsilon''' - \varepsilon''),
    \end{align}
    where the equality follows from the definition of $\tilde d$, and the inequality follows from the PAC bound on $\tilde\lambda$.
\end{proof}

\begin{proof}[Proof of Theorem~\ref{thrm:connect}]
    With the choice of the parameters in the theorem, we have
    \begin{align}
        \theta - (\varepsilon'-\varepsilon''') & = \theta \\
        (\rho - (N-1)\,(\varepsilon''-\varepsilon'''')) & = \rho \\
        (1-\delta')^{1/(N+1)} & = 1 - \delta.
    \end{align}
    Hence, Lemma~\ref{lemma:connect} guarantees that the constant pivot rule $h_n(t_{-n})=\tilde\kappa_n(\theta)-\tilde d_n$ satisfies DSIC, DE, $\theta$-IR, and $\rho$-WBB with probability $1-\delta$.
\end{proof}

\begin{proof}[Proof of Proposition~\ref{prop:connect}]
    The constant pivot rule in Theorem~\ref{thrm:connect} can be learned with $N$ independent runs of an $(\varepsilon,\delta')$-PAC BME and a single run of an $(\varepsilon,\delta')$-PAC estimator for an expectation, whose overall sample complexity is $O(N\,(K/\varepsilon^2)\,\log(1/\delta'))$.  The proposition can then be established by substituting $\delta'=1-(1-\delta)^{1/(N+1)}$.
\end{proof}

\subsection{Proofs of the propositions in Appendix~\ref{sec:detail:exp:moderate}}

\begin{proof}[Proof of Proposition~\ref{prop:se-bme}]
    Let $\mu_\star\coloneqq\max_{k\in[1,K]}\mu_k$ be the best mean, $\hat\mu$ be the best mean estimated by Algorithm~\ref{alg:SE-BME}, and $\hat\mu_k^{(t)}$ be the average of the first $t$ samples from arm $k$.
        Let $\alpha_t
        \coloneqq
        \sqrt{\frac{1}{2\,t}\log\left(\frac{\pi^2\,K\,t^2}{3\,\delta}\right)}$.
    Then we have
    \begin{align}
        \lefteqn{\Pr(|\hat\mu-\mu_\star|\le\varepsilon)} \notag\\
        & \ge \Pr(\mbox{At every iteration, the error in the estimated mean is less than $\alpha_t$ for any arm in $\cR$.\footnotemark})\\
        & = \Pr\left(
        \bigcap_{t=1}^\infty
        \bigcap_{k\in[1,K]}
        \{|\{\hat\mu_k^{(t)} - \mu_k| < \alpha_t\}
        \right)\\
        & = 1 - \Pr\left(
        \bigcup_{t=1}^\infty
        \bigcup_{k\in[1,K]}
        \{|\{\hat\mu_k^{(t)} - \mu_k| \ge \alpha_t\}
        \right)\\
        & \ge 1 - \sum_{t=1}^\infty
        \sum_{k\in[1,K]} \Pr\left(
        \{|\{\hat\mu_k^{(t)} - \mu_k| \ge \alpha_t\}
        \right)
        \quad\mbox{by union bound}\\
        & \ge 1 - 2 \, K \, \sum_{t=1}^\infty
        \exp\left( - 2 \, \alpha_t^2 \, t\right)
        \quad\mbox{by Hoeffding's inequality}\\
        & = 1 - \delta \, \frac{6}{\pi^2} \, \sum_{t=1}^\infty \frac{1}{t^2}
        \quad\mbox{by definition of $\alpha_t$}\\
        & = 1 - \delta.
    \end{align}
    \footnotetext{This condition suffices because it ensures that the best arms always remain in $\cR$.}
\end{proof}

\begin{proof}[Proof of Proposition~\ref{prop:se-bai}]
Let $\alpha_t\coloneqq\sqrt{\log\left(\frac{\pi^2\,K\,t^2}{6\,\delta}\right) / (2\,t)}$.  Let $\cB$ be the set of the (strictly) best arms.  Let $\hat\mu_k^{(t)}$ be the average of the first $t$ samples from arm $k$.  Then we have
\begin{align}
  \lefteqn{\Pr(\mbox{Algorithm~\ref{alg:SE-BAI} selects an $\varepsilon$-best arm.})} \notag\\
    & \ge \Pr(\mbox{At every iteration, all arms in $\cB$ remain in $\cR$ and any arm in $\cR$ is $2\alpha_t$-best.\footnotemark}) \\
    & \ge \Pr\left(\bigcap_{t=1}^\infty
    \bigcap_{k\in\cB} \{\hat\mu_k^{(t)} > \mu_k - \alpha_t\}
    \bigcap_{\ell\not\in\cB} \{\hat\mu_\ell^{(t)} < \mu_\ell + \alpha_t\}
    \right)\\
    & = 1 - \Pr\left(\bigcup_{i=1}^\infty
    \bigcup_{k\in\cB} \{\hat\mu_k^{(t)} \le \mu_k - \alpha_t\}
    \bigcup_{\ell\not\in\cB} \{\hat\mu_\ell^{(t)} \ge \mu_\ell + \alpha_t\}
    \right)\\
    & \ge 1 - \sum_{t=1}^\infty
    \left(
    \sum_{k\in\cB} \Pr(\hat\mu_k^{(t)} \le \mu_k - \alpha_t)
    +
    \sum_{\ell\not\in\cB} \Pr(\hat\mu_\ell^{(t)} \ge \mu_\ell + \alpha_t)
    \right)
    \quad\mbox{by union bound}\\
    & \ge 1 - \sum_{t=1}^\infty K \, \exp(-2\,\alpha_t^2\,t)
    \quad\mbox{by Hoeffding's inequality}\\
    & = 1 - \delta \, \frac{6}{\pi^2} \, \sum_{t=1}^\infty \frac{1}{t^2}
    \quad\mbox{by the definition of $\alpha_t$}\\
    & = 1 - \delta.
\end{align}
\footnotetext{This condition suffices because the algorithm stops either when $|\cR|=1$ or when $\alpha_t\le\varepsilon/2$, which implies that only $\varepsilon$-best arms are in $\cR$ when the algorithm stops.}
\end{proof}

\end{document}